\documentclass[review]{elsarticle}
\usepackage{float}
\usepackage{multirow}
\usepackage{amsthm}
\usepackage{graphicx}
\usepackage[export]{adjustbox}
\usepackage{changepage}
\usepackage{url}
\usepackage{amssymb}
\usepackage{ulem}
\usepackage{array}
\usepackage{booktabs}
\usepackage[table]{xcolor}
\usepackage[nointegrals]{wasysym} % symbols
\usepackage{listings}
\usepackage{pdflscape}
\usepackage{rotating}
\usepackage[utf8]{inputenc}
\usepackage{tikz,lipsum,lmodern}
\usepackage[most]{tcolorbox}
\usepackage{pgfplots}
\usepackage[toc,page]{appendix}
\usepackage{textcomp}
\usepackage{epstopdf}
\usepackage{csquotes}
\usepackage{dirtytalk}
\usepackage{parskip}
\usepackage{amssymb}
\usepackage{tikz}
\usepackage{comment}
\renewcommand{\arraystretch}{1.5} %adding space between rows in table
\usepackage{indentfirst}% this helps to indent the first paragraph after the heading or chapter
\usepackage[colorinlistoftodos,prependcaption,textsize=tiny]{todonotes}
\usepackage[utf8]{inputenc}
\usepackage[english]{babel}
\usepackage{geometry}
\pgfplotsset{compat=1.9}
\usepackage{lineno}
\usepackage{caption} 
\usepackage{adjustbox}
\usepackage{graphicx}
\usepackage{multirow}
\usepackage{pdfpages}
\usepackage{rotating}
\usepackage{ragged2e} 
\usepackage{xurl}
\usepackage[T1]{fontenc}
\usepackage{amsmath}
\usepackage{xspace}
\usepackage{hyperref}
\usepackage{fontawesome} 
\usepackage{longtable,tabu}
\usepackage{afterpage}
\usepackage{verbatim}
\usepackage{subfigure}
\usepackage{subfloat}
\usepackage{longtable}

\hypersetup
{
  colorlinks   = true, %Colours links instead of ugly boxes
  urlcolor     = red,  %Colour for external hyperlinks
  linkcolor    = blue, %Colour of internal links
  citecolor    = blue  %Colour of citations
}
\journal{Information and Software Technology}

\begin{document}
\begin{sloppypar}
\begin{frontmatter}

\title{Architecture Decisions in Quantum Software Systems: An Empirical Study on Stack Exchange and GitHub}




\author[mymainaddress]{Mst Shamima Aktar}
\ead{shamima@whu.edu.cn}
\address[mymainaddress]{School of Computer Science, Wuhan University, 430072 Wuhan, China}

\author[mymainaddress]{Peng Liang\corref{mycorrespondingauthor}}
\cortext[mycorrespondingauthor]{Corresponding author at: School of Computer Science, Wuhan University, China. Tel.: +86 27 68776137; fax: +86 27 68776027.}
\ead{liangp@whu.edu.cn}

\author[mysecondaryaddress]{Muhammad Waseem}
\ead{muhammad.m.waseem@jyu.fi}
\address[mysecondaryaddress]{Faculty of Information Technology, University of Jyväskylä, FI-40014 Jyväskylä, Finland}

\author[mythirdaddress]{Amjed Tahir}
\ead{a.tahir@massey.ac.nz}
\address[mythirdaddress]{School of Mathematical and Computational Sciences, Massey University, 4442 Palmerston North, New Zealand}

\author[myfourthdaddress]{\\Aakash Ahmad}
\ead{a.ahmad13@lancaster.ac.uk}
\address[myfourthdaddress]{School of Computing and Communications, Lancaster University Leipzig, 04109 Leipzig, Germany}

\author[mymainaddress]{Beiqi Zhang}
\ead{zhangbeiqi@whu.edu.cn}

\author[myfifthaddress]{Zengyang Li}
\ead{zengyangli@ccnu.edu.cn}
\address[myfifthaddress]{School of Computer Science, Central China Normal University, Wuhan, China}

\begin{abstract}
\justifying
\textbf{Context}: Quantum computing provides a new dimension in computation, utilizing the principles of quantum mechanics to potentially solve complex problems that are currently intractable for classical computers. However, little research has been conducted about the architecture decisions made in quantum software development, which have a significant influence on the functionality, performance, scalability, and reliability of these systems.

\noindent\textbf{Objective}: The study aims to empirically investigate and analyze architecture decisions made during the development of quantum software systems, identifying prevalent challenges and limitations by using the posts and issues from Stack Exchange and GitHub.

\noindent\textbf{Method}: We used a qualitative approach to analyze the obtained data from Stack Exchange Sites and GitHub projects - two prominent platforms in the software development community. Specifically, we collected data from 385 issues (from 87 GitHub projects) and 70 posts (from 3 Stack Exchange sites) related to architecture decisions in quantum software development. 

\noindent\textbf{Results}: The results show that in quantum software development (1) architecture decisions are articulated in six linguistic patterns, the most common of which are \textit{Solution Proposal} and \textit{Information Giving}, (2) the two major categories of architectural decisions are \textit{Implementation Decision} and \textit{Technology Decision}, (3) \textit{Software Development Tools} are the most common application domain among the twenty application domains identified, (4) \textit{Maintainability} is the most frequently considered quality attribute, and (5) \textit{Design Issues} and \textit{High Error Rates} are the major limitations and challenges that practitioners face when making architecture decisions in quantum software development.

\noindent\textbf{Conclusions}: Our results show that the limitations and challenges encountered in architecture decision-making during the development of quantum software systems are strongly linked to the particular features (e.g., quantum entanglement, superposition, and decoherence) of those systems. These issues mostly pertain to technical aspects and need appropriate measures to address them effectively. 

\end{abstract}


\end{frontmatter}

\section{Introduction} \label{introduction} 
Quantum computing is a rapidly developing technology that is a combination of physics, chemistry, mathematics, computer science, and information theory to revolutionize computation and problem-solving capabilities \cite{marella2020introduction}. Classical computing processes information using classical bits, which can represent a value of either 0 or 1. In contrast, quantum computing introduces qubits or quantum bits, which can represent two-state $|0\rangle$ and $|1\rangle$, serving as the fundamental units of quantum information technology \cite{steffen2011quantum}. As a mature technology, classical computing is making significant contributions to the advancement of today's digital world \cite{lloyd2000ultimate}\cite{mavroeidis2018impact}. Despite its invaluable contribution to the landscape of computation, classical computing faces inherent limitations in data encoding, processing, and transmission, particularly when managing complex tasks such as optimizing energy systems, conducting large-scale simulations, and enhancing cryptographic security areas - where quantum computing shows enormous potential \cite{gruska2001quantum}\cite{ajagekar2019quantum}\cite{vyakaranal2018performance}\cite{li2024quantum}. As a result, quantum computing has largely been motivated by academic (e.g., institutions like MIT and Stanford doing research in quantum algorithms) and industrial interests (e.g., companies like Google and IBM developing quantum technologies) to change the scenery of information technology, and the role of quantum software architecture has been recognized to bring impactful changes in this transformation process \cite{srivastava2016commercial}\cite{svore2004toward}\cite{bhasin2021quantum}. Integrating quantum software architecture and programming with quantum hardware is a foundational aspect of quantum computing research \cite{9590459}\cite{mccaskey2018language}. This integration makes it possible to develop and execute quantum algorithms designed to utilize quantum bit's power and solve complex problems more efficiently than classical algorithms \cite{preskill2018quantum}. The most fundamental difference between quantum and classical algorithms lies in their state representation - classical algorithms use probabilities while quantum algorithms use amplitudes \cite{strubell2011introduction}. Therefore, central to the potency of quantum computing are quantum algorithms, including Shor's large numbers factoring algorithm \cite{willsch2023large} and Grover's quantum search algorithm \cite{mandviwalla2018implementing}. Quantum algorithms are used across various domains,  such as cryptography, search and optimization processes, and solving linear equation systems \cite{montanaro2016quantum}\cite{harrow2009quantum}. Quantum programming entails developing algorithms specifically for quantum computers and reducing the execution costs of quantum algorithms on future devices \cite{heim2020quantum}. Quantum programs are written using domain-specific programming languages such as Q\# and Silq, and quantum software architecture is important to develop quantum software systems \cite{zhao2020quantum}\cite{sundaram2022quantum}. However, the fields of quantum software architecture and quantum software engineering are still in their early stages of development \cite{dey2020liqui}\cite{behera2019designing}\cite{courtland2017google}.  Compounding this, previous studies highlight the considerable obstacles that quantum software development grapples with, such as intricate programming paradigms, scarce simulation resources, high error frequencies, a dearth of efficient debugging tools, and limited software libraries. These obstacles accentuate the intricacies involved in architecture decisions for quantum software systems \cite{7927104}\cite{ali2022software}.

In the domain of software, quantum computing is ready to begin renovations ``golden age'', changing the landscape of software engineering \cite{piattini2021quantum}. Over the last few years, the development of quantum programming languages, algorithms, libraries, and tools has received much attention \cite{serrano2022quantum}, quantum software architecture has so far received little attention, which significantly impacts the research and practice of quantum software systems \cite{yue2023challenges}. Different companies (e.g., Qiskit\footnote{\url{https://qiskit.org/}} from IBM, Cirq\footnote{\url{https://quantumai.google/cirq}} from Google Research, Q\#\footnote{\url{https://learn.microsoft.com/en-us/azure/quantum}} from Microsoft, and C++\footnote{\url{https://www.intel.com/content/www/us/en/developer/tools/quantum-sdk/overview.html}} from Intel) and cloud providers (e.g., Amazon Web Services, Alibaba Cloud, and Azure Quantum) have already invested in tools as well as programming languages that allow developers, researchers, and users to solve real-world problems (e.g., risk assessment and fraud detection in banking, drug discovery, protein folding, supply chain optimization, energy distribution, and advertising scheduling) \cite{kommadi2020quantum}. Quantum programming languages emphasize computational and implementation specifics to create executable specifications, yet often fail to incorporate a comprehensive, global perspective on the software systems being designed can lead to a reduction in essential architecture views, potentially compromising the quality and functionality of the final quantum software product \cite{zhao2020quantum}\cite{moguel2020roadmap}\cite{piattini2021toward}. Concurrently, quantum computing has an impact on all software life cycle processes and techniques, and quantum software encounters multifaceted challenges \cite{garcia2023quantum}, such as making existing software quantum-safe \cite{zhang2023making}.

Software architecture is the high-level design and organization of a software system, including its structural components, interactions, and behavior \cite{bass2012software}. Software architecture for quantum software systems structures quantum-intensive systems, utilizing quantum bits and quantum gates as the essential components and connectors for the software architecture design and implementation \cite{zhao2024unraveling}\cite{khan2023software}. The architecture of a quantum software system is designed through a decision-making process, where the architecture is viewed as a composition of a set of explicit design decisions \cite{el2023secure}\cite{jansen2005software}. Decision-making significantly influences the success or failure of a software system, as a high-quality architecture decision-making process can lead to desirable results, while a poor-quality architecture and suboptimal architecture decisions more likely lead to undesirable outcomes and even architectural technical debt \cite{malavolta2014enhancing}\cite{li2015architectural}. Making architecture decisions during the development process of quantum software systems is complex, requiring a comparative structure to analyze and select among various tools, services, and techniques based on specific use cases and algorithms \cite{vietz2021decision}. Architecture decisions are increasingly influenced by practitioner feedback, focusing on hybrid quantum-classical patterns and service orientation to efficiently address emerging computational challenges in quantum software \cite{ahmad2022towards}. Architecture decisions also affect many aspects of a software system, including its quality (e.g., maintainability, testability, and reliability) of quantum software development \cite{sodhi2021quantum}. Recent workshops, such as the Quantum Software Architecture Workshop (QSA), highlight the growing importance of architectural considerations in the domain of quantum computing \cite{barzen20222nd}. Despite the essential importance of architecture decisions in the development of quantum software systems, architecture decisions are unexplored in the area of quantum software development \cite{sodhi2021quantum}. Software architects encounter various quantum software architecture limitations and challenges and should carefully consider overcoming architecture challenges in a quantum software system \cite{khan2023software}\cite{yue2023challenges}. Recent studies also identify complex challenges, particularly in quantum software development, which is directed towards the intricate process of architecting and implementing quantum software systems, with structural difficulties surfacing even for seasoned architects and developers \cite{ahmad2022towards}.  To tackle the challenges of quantum software architecture, the current generation of architects and developers often find themselves inadequately prepared for the development of quantum software systems \cite{naveh2021quantum}\cite{awan2022quantum}\cite{polian2015design}. Despite the myriad of recent literature on quantum software engineering and architecture, a noticeable gap remains with the lack of empirical studies analyzing critical architecture decision-making within quantum software systems \cite{yue2023challenges}\cite{zhao2024unraveling}\cite{murali2019full}. To this end, this study aims to analyze the architecture decisions as well as the challenges in architecture decision-making associated with quantum software development from discussions made between developers in GitHub issues and Stack Exchange posts. 

To achieve the goal of this study (see Section \ref{goalandRQs}), we conducted an empirical study to investigate architectural decisions made in quantum software development. We collected data from two sources: GitHub issues (of open-source quantum software projects) and Stack Exchange Q\&A sites (Stack Overflow, Quantum Computing, and Computer Science). More specifically, we collected 3,185 open and closed issues (from 192 GitHub projects) and 1494 posts (from the Stack Exchange sites). We manually filtered out irrelevant issues and posts, and we finally got 385 related issues (from 87 quantum software projects) and 70 posts from Stack Exchange. Furthermore, we manually extracted data from the 385 related issues and 70 posts and analyzed the extracted data using a predefined classification \cite{7371991} and the Constant Comparison method \cite{grove1988analysis} for answering the research questions.

\textbf{Our study results show that} (1)\textit{Solution Proposal} and \textit{Information Giving} are the most frequently used linguistic patterns, (2) \textit{Implementation Decision} and \textit{Technology Decision} are the two main types of architecture decisions, (3) \textit{Software Development Tool} is the most common application domain identified, (4) \textit{Maintainability} is the dominant quality attribute considered, and (5) \textit{Design Issues} and \textit{High Error Rates} are the main limitations and challenges when making architecture decisions in quantum software development.

\textbf{The contributions of this work:} (1) we identified the linguistic patterns practitioners utilized when making architecture decisions in developing quantum software systems; (2) we investigated several facets of architecture decisions in quantum software development, such as the categories of decisions made, the application domains, and the quality attributes considered; (3) we present and discuss the challenges and limitations developers face while designing quantum software systems; (4) we provide guidelines to help practitioners design quantum software systems with informed architectural decisions; and (5) we constructed a dataset \cite{dataset} that collects the architecture decisions made in quantum software development.

The rest of this paper is structured as follows: Section \ref{Background} presents the background of the study. Section \ref{researchdesign} describes the research methodology. Section \ref{results} provides the study results, followed by a discussion of these results in Section \ref{discussionAndImplications}. Section \ref{ThreatValidity} clarifies the validity threats. Section \ref{relatedwork} discusses the related work. Finally, Section \ref{conclusionFurtureWork} concludes the work with future research directions.

\section{Background}\label{Background}

\begin{figure}[h]
 \centering
 \includegraphics[width=1.0\linewidth]{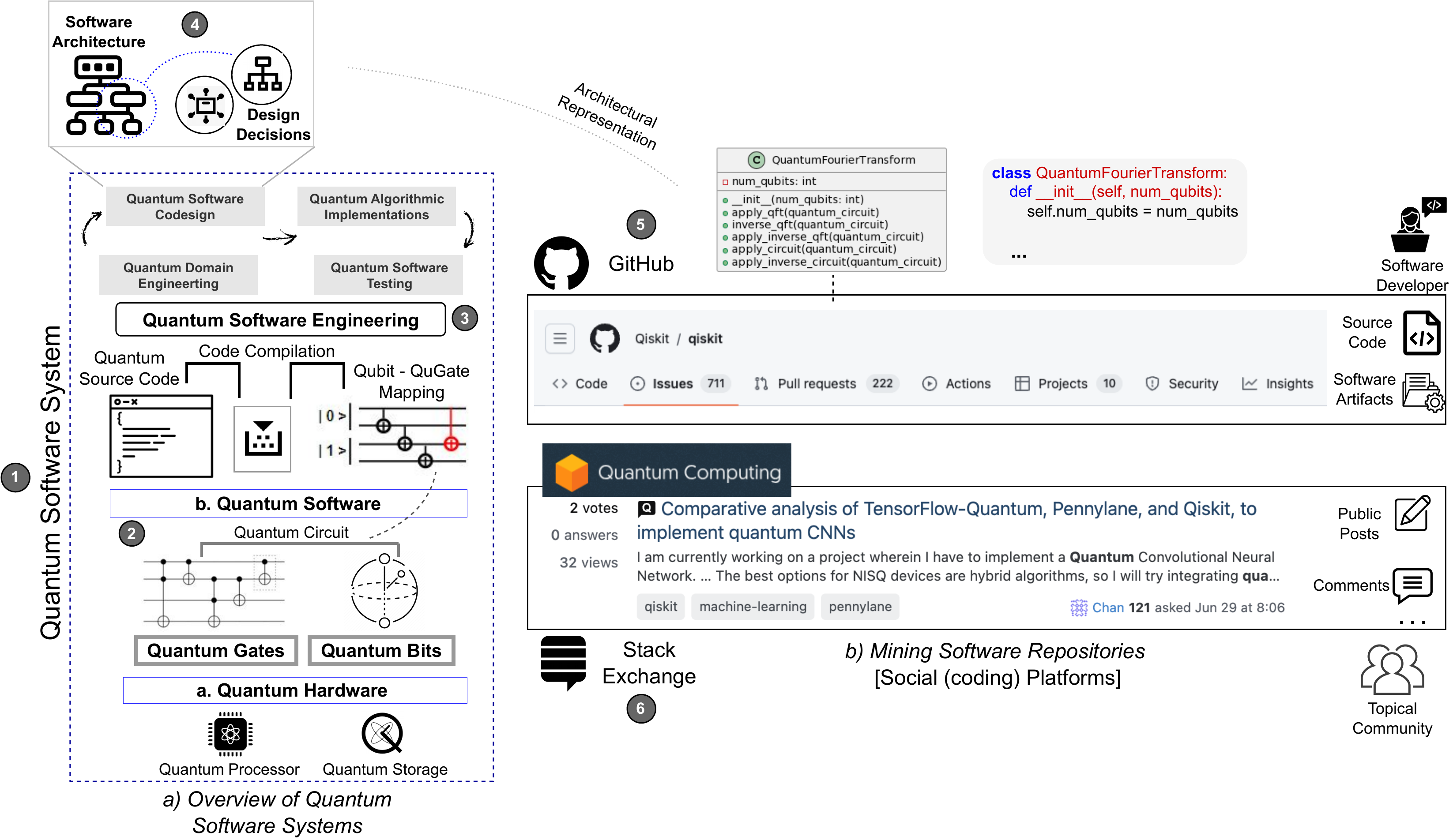}
 \caption{ a) Overview of Quantum Computing Systems, b) Mining Software Repositories
[Social (coding) Platforms] }
 \label{BackgroundFigure}
\end{figure}

In this section, we introduce the background concepts used in this study, including quantum computing, quantum software engineering, and architecture decisions. The overview of the background is shown in Fig. \ref{BackgroundFigure}.

\subsection{Quantum computing}
Quantum computing is a model of computation that focuses on designing and developing computer-based systems utilizing the principles of quantum mechanics (e.g., superposition, entanglement, and interference) to process information.
\begin{itemize}
\item \textbf{Superposition}: Superposition is a fundamental principle of quantum mechanics that allows quantum systems, or quantum bits (qubits), to exist simultaneously in multiple states. Unlike classical bits that can only be 0 or 1 at any one time, a qubit can occupy the states $|0\rangle$ and $|1\rangle$, or any quantum superposition of these states. This superposition capability represents qubit states within a matrix format, as illustrated in Equation \ref{equation1}. This capability allows qubits to perform parallel computations, enhancing computational efficiency and optimizing system performance. 
\begin{equation}\label{equation1}
|0\rangle = 
\begin{bmatrix}
1 \\
0 \\
\end{bmatrix},
\quad
|1\rangle = 
\begin{bmatrix}
0 \\
1 \\
\end{bmatrix}
\end{equation}
\item \textbf{Entanglement}: Quantum entanglement is a unique feature of quantum physics, which is not present in classical mechanics. Quantum entanglement uniquely connects quantum bits (qubits), so the state of one qubit instantly correlates with the state of another, regardless of the distance between them. Consequently, researchers cannot describe the state of each qubit independently of the others. Entangled qubits can thus provide a robust synchronization necessary for quantum computing, which is essential for specific algorithms that require complex correlation patterns.
\item \textbf{Interference}: Qubits, the fundamental units of quantum information, can exist in multiple states simultaneously due to superposition. This characteristic leads to quantum interference, which reflects the likelihood of qubits assuming specific states upon measurement. Quantum computing systems, therefore, implement strategies to minimize this interference, aiming to enhance the precision of their computational outcomes.
%\item \textbf{Decoherence}: Decoherence involves the loss of quantum coherence, whereby qubits interacting with their environment lose their quantum properties and revert to classical states. This interaction poses significant challenges in maintaining stable and reliable quantum computations. Techniques for quantum error correction and isolation are critical to preserve the coherent states necessary for computation.
\end{itemize}

Quantum gates are fundamental to quantum computing, analogous to classical logic gates but designed for operations on qubits. These gates manipulate qubits in superposition states and are typically represented as matrices that execute specific operations on targeted qubits. The most frequently encountered gates include the Pauli X (Eq. \ref{equation2}), Pauli Y (Eq. \ref{equation3}), Phase Shift (Pauli Z) gate (Eq. \ref{equation4}), and the Hadamard gate (Eq. \ref{equation5}). These gates play a foundational role in quantum mechanics, often referred to as the `quantum state basis', translating algorithmic mathematics into quantum circuits, assessing state probabilities, and forecasting an algorithm's performance. For a qubit $|X\rangle$ to be in this state basis, it must conform to a linear combination represented by $|x\rangle = \alpha |0\rangle + \beta |1\rangle$, where \( \alpha \) and \( \beta \) are complex numbers ensuring the normalization condition equals 1 that is,  \( |\alpha|^2 + |\beta|^2 = 1 \) This concept is also known as the superposition of two basis states.
\begin{equation}\label{equation2}
X = 
\begin{bmatrix}
0 & 1 \\
1 & 0 \\
\end{bmatrix}
\end{equation}
\begin{equation}\label{equation3}
Y = 
\begin{bmatrix}
0 & -\mathrm{i} \\
\mathrm{i} & 0 \\
\end{bmatrix}
\end{equation}
\begin{equation}\label{equation4}
Z = 
\begin{bmatrix}
1 & 0 \\
0 & -1 \\
\end{bmatrix}
\end{equation}
\begin{equation}\label{equation5}
H = \frac{1}{\sqrt{2}}
\begin{bmatrix}
1 & 1 \\
1 & -1 \\
\end{bmatrix}
\end{equation}

In the early 1980s, physicist Richard Feynman proposed the notion of simulating quantum systems using quantum mechanical processes \cite{feynman2018simulating}. Over a couple of decades, quantum computing has transitioned from theoretical concepts to practical implementations, driven by exponentially increasing computational potential \cite{sofge2008survey}. This transformation has motivated research into quantum algorithms, error correction, and quantum simulation~\cite{childs2018toward}. Quantum computing plays an important role in several fields of computer science most notably natural language processing \cite{coecke2020foundations}, machine learning \cite{biamonte2017quantum}, and cyber security \cite{9130525}. To realize practical quantum applications, it is essential to have quantum software tools for various design tasks across multiple abstraction levels, along with practical benchmarks to empirically evaluate and compare these tools against the current state of the art \cite{quetschlich2023mqt}. However, the development of practical quantum systems remains difficult due to the lack of reliability requirements of quantum computing algorithms, physical machines foreseen, and hardware limitations \cite{chong2017programming}. Quantum Computing as a Service (QCaaS) by establishing a reference architecture, demonstrates its practical application through microservices in the quantum-classic divide and validates its effectiveness and reusability via insights from professionals in the field of quantum software engineering. 

%From the 1980s, the concept of quantum computing to today's reality of performing calculations on actual quantum computers, and this rapid technological advancement presents new possibilities for integrating quantum technology into existing software \cite{grossi2021serverless}. At first, physicist Richard Feynman proposed the notion of simulating quantum systems using quantum mechanical processes \cite{feynman2018simulating}. Over a couple of decades, quantum computing has transitioned from theoretical concepts to practical implementations, driven by exponentially increasing computational potential \cite{sofge2008survey}. To realize practical quantum applications, it is essential to have quantum software tools for various design tasks across multiple abstraction levels, along with practical benchmarks to empirically evaluate and compare these tools against the current state of the art \cite{quetschlich2023mqt}. However, the development of practical quantum systems remains difficult due to the lack of reliability requirements of quantum computing algorithms, physical machines foreseen, and hardware limitations \cite{chong2017programming}. Quantum Computing as a Service (QCaaS) by establishing a reference architecture, demonstrates its practical application through microservices in the quantum-classic divide and validates its effectiveness and reusability via insights from professionals in the field of quantum software engineering \cite{ahmad2023reference}.

\subsection{Quantum software engineering}
Quantum Software Engineering (QSE) represents a transformative approach combining the principles of quantum mechanics with the established practices of software engineering, aiming to design, develop, validate, and evolve software systems and applications fit for the quantum era \cite{ahmad2022towards}. The transition from theoretical concepts to practical implementations in quantum computing has seen significant milestones, such as Google's demonstration of ``quantum supremacy'' in October 2019, showcasing quantum computers solving problems beyond the reach of classical computers \cite{bobier2021happens}. Classical software engineering involves methods and techniques for developing large-scale software systems, using the engineering metaphor to highlight a systematic approach that meets organizational requirements and constraints \cite{van2008software}. On the other hand, QSE signifies a new era of software-intensive engineering activities designed to develop applications that manage the underlying hardware, enabling the cost-effective and scalable development of dependable quantum software systems across various domains \cite{ali2022software}\cite{piattini2021toward}. Compared with classical software engineering, quantum software engineering has inherent complexity and presents multifaceted challenges  (e.g., the requisite for physics knowledge among developers who often lack software engineering experience, particularly software processes, quality practices,  complex quantum algorithms, and programming)  that emphasize the significance of QSE \cite{felderer2023software}. The life cycle in QSE is comprehensive, encompassing quantum requirements engineering, design, implementation, testing, and maintenance, requiring collaborative efforts from both scientific and industry to propel real-world applicability  \cite{akbar2022quantum}. The intersection of software engineering and quantum software engineering underscores the necessity for a tailored software development life cycle, which accentuates the importance of requirements engineering, particularly in the quantum domain, where it facilitates communication among stakeholders and helps in comprehending complex application domains \cite{dwivedi2024quantum}. QSE is a growing field of study and Quantum Software Architectures (QSA) is a sub-domain of study - both are understudied and face research challenges \cite{yue2023challenges}. More specifically, QSE provides the methodologies, techniques, and tools required to systematically develop, operate, and maintain software for quantum computers, while QSA defines the structure of the system, including the quantum software components, their properties, and the interfaces and interactions between them that are involved in the design of large-scale quantum software applications and their implementation on quantum hardware. This task is challenging, requiring software development that aligns with the laws of quantum comprehension \cite{DESTEFANO2022111326}\cite{wecker2014liqui}. QSA is often characterized by unique challenges such as defining suitable quantum algorithms, handling quantum data structures, and the crucial task of choosing suitable quantum hardware for the optimal execution of these software architectures, which are part of the larger research hurdles within QSE \cite{9590459}\cite{moguel2020roadmap}. Both the software engineering and software architecture research communities have been making concerted efforts to cultivate dedicated platforms such as conferences and workshops. The aim is to define target outcomes, simplify emerging issues, and introduce collective initiatives pertaining to the engineering and architecture planning for the development and structural design of quantum software \cite{abreu2021first}\cite{barzen20222nd}\cite{murillo2024challenges}. Recent studies emphasize several particularly difficult aspects of QSE that diverge from traditional software; including requirements issues and challenges for architecting, interpreting quantum program outputs,  explaining the theory behind quantum computing code, reducing the understanding gap between quantum and classical computing, and testing QSE applications, along with pinpointing opportunities that arise from these very complexities \cite{felderer2023software}\cite{li2021understanding}. QSE seeks to deal with those evolving challenges of quantum software development, requiring particular architecture decisions within the decision-making process to overcome these challenges \cite{akbar2023systematic}.

\subsection{Architecture decision}
 Software architecture is a blueprint that guides the development process, ensuring the system's quality attributes (e.g., maintainability, performance, scalability) and aligning technical decisions with business goals \cite{ducasse2009software}. Classical software architecture is the set of structures of a software system that consists of the various components, the relationships among those components, and their respective properties, which are necessary to understand and create the system \cite{bass2012software}. Quantum software architecture represents a novel category of software architecture that can abstract complicated and implementation-specific activities while offering architectural descriptions (e.g., components, connections, and configurations) to design and develop quantum software \cite{khan2023software}. In the realm of software engineering and architecture design, software architecture decisions can be defined as critical design decisions that address significant architecture requirements and are generally considered complex to formulate and potentially expensive to alter \cite{van2016decision}. A comparative analysis reveals that architecture decisions in classical software development focus on optimizing performance, scalability, and maintainability using deterministic algorithms and well-defined architecture patterns and binary data structures \cite{ramirez2016comparative}, while architecture decisions in quantum software development focus on quantum algorithms and well-defined quantum architecture patterns and quantum data structures to address unique challenges (e.g., algorithms, programming, design, implementation, and testing), which require suitable architecture decisions \cite{selva2016patterns}\cite{nallamothula2020selection}. The importance of architecture decisions in quantum software systems is also highlighted by a systematic review on software architecture for quantum computing systems \cite{khan2023software}. In many contexts, including the Internet of Things (IoT) \cite{alreshidi2019architecting}, machine learning-based systems \cite{muccini2021software}, and microservices \cite{waseem2022decision}, empirical studies and systematic mappings emphasize the importance of architecture decisions. The documentation, communication, and evaluation of these crucial decisions are made easier by existing models and tools \cite{shahin2009architectural}. As new architectural styles and paradigms arise, these decisions help developers manage common challenges in architecture design \cite{jansen2005software}. The relevance of architecture decisions even in cutting-edge fields is reflected in practical perspectives on agile quantum software development \cite{khan2022agile}. When software systems become more complex, understanding, clearly expressing, and making effective architecture decisions are essential for achieving software quality and meeting stakeholder needs.

\section {Research design} \label{researchdesign}

\subsection{Research goal and research questions} \label{goalandRQs}

\begin{figure}[h]
\centering
\includegraphics[width=1.0\linewidth]{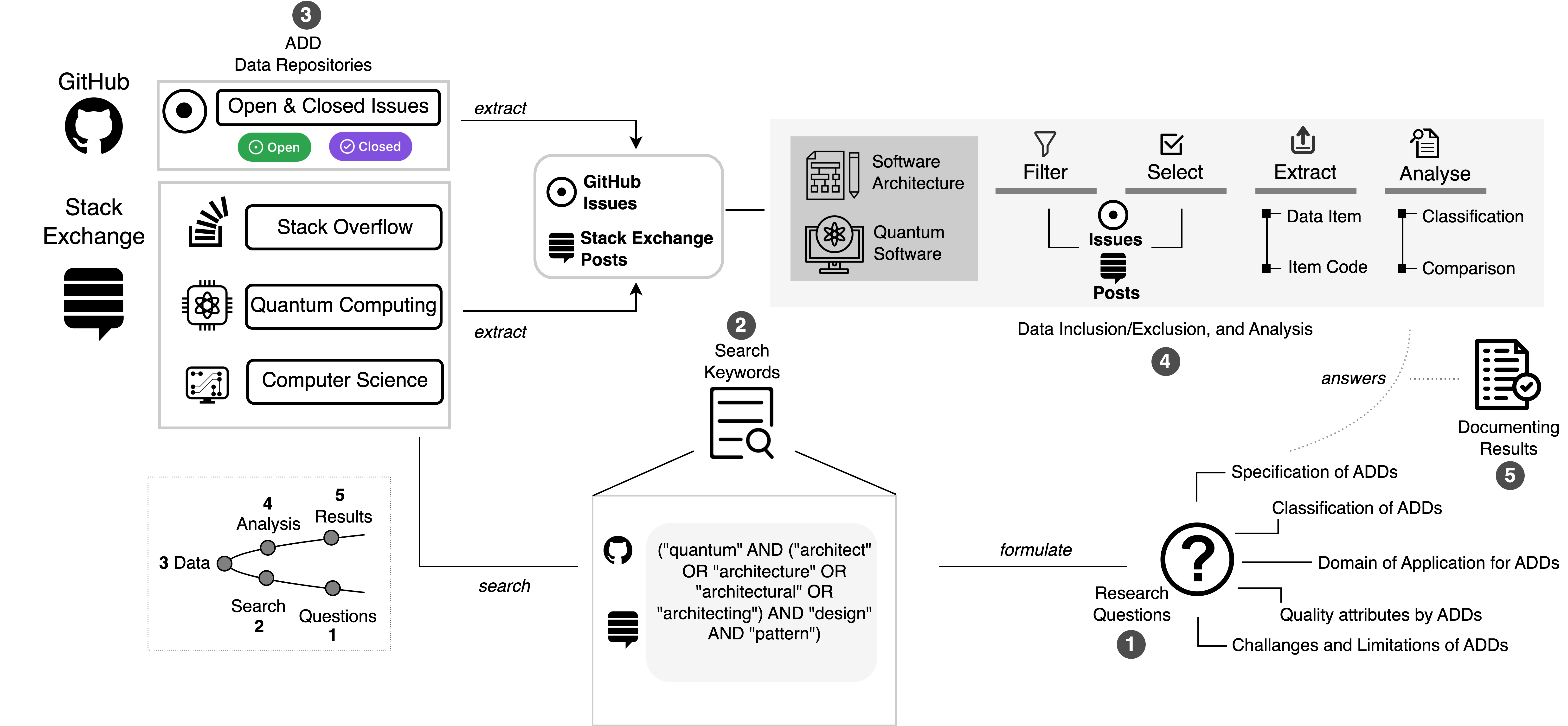}
\caption{Overview of the research process}
\label{OverviewOfTheResearchProcess}
\end{figure}

The goal of this study is formulated using a Goal-Question-Metric approach \cite{caldiera1994goal}: We aim to {\textit{\textbf{analyze} architecture decisions made during the development of quantum software \textbf{for the purpose of} investigating their decision description and categorization, application domains covered, quality attributes considered, and the limitations and challenges \textbf{from the point of view of} practitioners \textbf{in the context of} architecture-related posts and GitHub issues of quantum software systems}}. In order to perform this empirical study, we adhered to the recommendations for conducting empirical studies by Easterbrook \textit{et al}. \cite{easterbrook2008selecting}. Following the defined goal, we provide five Research Questions (RQs), their rationale (see Table~\ref{table:researchQuestions}), and an overview of the research process (see Fig. \ref{OverviewOfTheResearchProcess}) in the subsections below.

%\subsection{Research Questions}\label{researchQuestionsSection} 

{\renewcommand{\arraystretch}{1}
\begin{table} [h!]
\small
\centering    
\caption{Research questions and their rationale} 
\label{table:researchQuestions}
\begin{tabular}{p{5cm}p{10.5cm}}
\toprule
\textbf{Research Question}& \textbf{Rationale}\\
\midrule

\textbf{RQ1.} How do developers express architecture decisions made in quantum software development? & This RQ aims to provide the linguistic patterns used by developers to express architecture decisions in the development of quantum software systems. Developers express architecture decisions in quantum software systems through various linguistic patterns to explain their ideas and communicate with their peers on different platforms (e.g., GitHub, Stack Exchange). The answer to this RQ can help developers identify architecture decision-related content in quantum systems development. \\ \hline
  
\textbf{RQ2.} What types of architecture decisions are made in quantum software development? & This RQ intends to categorize the architecture decisions made in quantum software systems. Developers make a number of architecture decisions while developing quantum software systems, and these decisions cover various aspects of architecture design (e.g., architecture patterns, and technology). The answer to this RQ can offer insights into the specific categories of architectural decisions being made in quantum software systems. \\ \hline
	 
\textbf{RQ3.} What are the application domains of quantum software in which architecture decisions are made? & This RQ aims to analyze the application domains of quantum software systems in which architecture decisions are made. Quantum technologies are widely employed in developing software systems across many application domains (e.g., optimization, cryptography, materials science, and machine learning). The answer to this RQ may provide guidance for making architectural decisions in particular domains. \\ \hline

\textbf{RQ4.} What quality attributes are considered when developers make architecture decisions in quantum software development? & This RQ aims to provide the quality attributes that developers focus on when making architecture decisions in quantum software development. Quantum software systems have dedicated concerns and the answer to this RQ can provide a comprehensive set of quality concerns highlighted in quantum software development. \\ \hline
	 
\textbf{RQ5.} What are the limitations and challenges of making architecture decisions in quantum software development?& This RQ aims to investigate the limitations and challenges developers may encounter when making architecture decisions for quantum software systems with related quantum components. Making architecture decisions in the creation of quantum software still has a number of limitations and challenges. The answer to this RQ can provide the directions for future exploration.
\\
\bottomrule 
\end{tabular}
\end{table}}

%We study domain-related which typically have one answer.We study domain-related which typically have one answer. We study architecture decision-related questions that can be more than one answer. We study domain-related which typically have one answer.  We study quality attributes-related questions that can be more than one answer.  We study how developers express architecture decisions that can be more than one answer. 

\subsection{Data collection} \label{datacollection}
We collected and analyzed data from 87 GitHub\footnote{\url{https://github.com}} projects and three Stack Exchange\footnote{\url{https://stackexchange.com/sites}} sites. GitHub hosts a large number of open-source projects and offers access to rich data from those projects, including version history and issue tracking data \cite{cosentino2017systematic}. Stack Exchange sites contain a number of technical Q\&A sites that cover various topics and domains, which include Stack Overflow\footnote{\url{https://stackoverflow.com/}}, one of the most active Q\&A sites that developers use from all over the world \cite{barua2014developers}. Quantum Computing Stack Exchange\footnote{\url{https://quantumcomputing.stackexchange.com/}} site is a dedicated platform for topics related to quantum computing. Computer Science Stack Exchange\footnote{\url{https://cs.stackexchange.com/}} is specifically tailored to the field of computer science questions. Many researchers have used GitHub and Stack Exchange sites for quantum software system research. For example, Li \textit{et al}. \cite{li2021understanding} explored the challenges related to Quantum Software Engineering (QSE) by analyzing data from GitHub and Stack Exchange sites. Their research focused on understanding the broader challenges perceived by developers in the area of QSE. Additionally, Openja \textit{et al}. \cite{openja2022technical} conducted an empirical analysis using GitHub data to explore the distribution and evolution of technical debt in quantum software, as well as their correlation with incidences of faults. Moreover, Zhang \textit{et al}. \cite{zhang2023architecture} empirically analyzed architecture decisions by collecting data from GitHub and Stack Overflow in the development of AI-based systems. However, our focus is to analyze architecture decisions in quantum software systems. Therefore, we selected specific GitHub projects and Stack Exchange discussions based on their relevance to quantum software development to ensure a rich dataset.

%Similarly, we decided to use both GitHub and Stack Exchange sites in our research. We specifically selected popular and active GitHub projects related to quantum software development to ensure a rich source of data and relevant discussions. Additionally, we chose Stack Exchange posts that are tagged with quantum computing topics to capture a broad spectrum of community interactions and expert insights [1][2]

\textbf{1) Data Collection from GitHub Projects}  

We followed a two-step process to collect quantum projects hosted on GitHub. First, we identified a set of quantum projects from GitHub. Then, we collect the issues from these projects. 

In the first step, we performed a pilot search with several terms before we decided on the most appropriate terms for selecting relevant quantum projects on GitHub, namely  ``\textit{quantum}'' and ``\textit{quantum computing}''. The search encompassed content within the projects  (e.g., name, topics, descriptions, and README files). We found 74,432 projects by using the keyword ``\textit{quantum}''. In contrast, using the keyword ``\textit{quantum computing}'' resulted in significantly fewer projects  (12,401), although with many irrelevant results such as books\footnote{\url{https://github.com/JackHidary/quantumcomputingbook}} and other learning resources\footnote{\url{https://github.com/desireevl/awesome-quantum-computing}}. Notably, ``\textit{quantum computing}'' is also a subdomain of the keyword ``\textit{quantum}''. During the search process by using the keyword ``\textit{quantum}'', we also discovered that the retrieved projects contained the keyword ``\textit{quantum computing}''. Therefore, we decided to use the keyword ``\textit{quantum}''. To mine those projects, we used GitHub REST API\footnote{\url{https://docs.github.com/en/rest?apiVersion=2022-11-28}}. We applied the following selection criteria: 1) projects that contain the word ``\textit{quantum}'' (i.e., including the name, description, topics, and README file); 2) the number of stars $\geq$ 50; and 3) the number of forks is $\geq$ 15 to minimize the possibility of including student assignments following the idea of prior studies \cite{waseem2021nature}. We identified 1,226 projects that meet the above selection criteria. Later, we used the following criteria to filter out our search results \cite{di2020topfilter}. We manually checked the repository (e.g., description, topics, and README file) and excluded those that pertained solely to quantum documentation or study materials and tutorials, not written in English, and repositories that are not related to quantum software systems but merely contain the word ``\textit{quantum}'' in their descriptions (e.g., FirefoxColor\footnote{\url{https://github.com/mozilla/FirefoxColor}}). In this step, 879 projects were removed, and 347 projects remained. 

%The term ``\textit{quantum computing}'' often draws attention to the theoretical projects.When we conducted searches using the ``\textit{quantum}'' keyword, we also retrieved projects containing ``\textit{quantum computing}'' keywords.

In the second step, we initially carried out a pilot search using various keywords that are the same as the architecture search terms ``\textit{architect*}'' (i.e., ``\textit{architect}'', ``\textit{architecture}'', ``\textit{architectural}'', ``\textit{architecting}'', ``\textit{architected}''), ``\textit{design*}'', and ``\textit{pattern*}'' and to search the title and body of the open and closed issues. We found issues that include the keywords (``\textit{architect}'', ``\textit{architecture}'', ``\textit{architectural}'', ``\textit{architecting}'', ``\textit{architected}'',``\textit{design}'', and ``\textit{pattern}''). Therefore, we decided to use those keywords and searched the title and body of the open and closed issues of the 347 quantum projects. After searching with the search terms, we found 3,185 open and closed issues containing the keywords, and those issues came from 192 quantum software projects. We ended up the data collection process from GitHub with 192 quantum software systems and 3,185 open and closed issues.

%we first used a keyboard-based search to locate repositories that contain the keyword ``\textit{quantum}'' (case insensitive) and noted the keyword was searched on the entire repository (i.e., including the name, description, topics, and README file). Initially, this keyword search returned a total of 73,506 repositories. In the second phase, we restricted our search criteria using the keyword ``\textit{quantum}''  (case insensitive), the number of stars $\geq$ 50, and the repository is mainline (i.e., not forked). The phase reduced the number of results to 1,577 repositories. In the third phase, following the idea of prior studies (\cite{waseem2021nature}), %we chose repositories that have been forked  $\geq$ 15 to minimize the possibility of including student assignments. In this step, 351 repositories were removed, and 1,226 repositories stayed.Finally, we searched the close issues (i.e., the title and body) to include architecture by the search terms i.e., ``\textit{architect*}'', ``\textit{architect}'', ``\textit{architecting}'', ``\textit{architected}'', ``\textit{architecture}'', ``\textit{architectural}'', and ``\textit{design}'' (note that we select those search keyword for getting architecture related issues). Finally, We got 165 quantum software systems with 1760 closed retrieved issues.

\renewcommand{\arraystretch}{1}
\begin{table} [h!]
\footnotesize % Reduce font size
\centering    
\caption{GitHub Projects, Number of Retrieved Issues, and Number of Related Issues} 
\label{gitHubProject}
\begin{tabular}{p{.6cm}p{2.3cm}p{1.2cm}p{1.2cm}p{0.6cm}|p{0.6cm}p{2.3cm}p{1.2cm}p{1.2cm}p{0.6cm}}
\toprule
\textbf{\#} & \textbf{GitHub Project}  & \textbf{Num of\newline Retrieved Issues}  & \textbf{Num of\newline Related Issues} & \textbf{Stars} & \textbf{\#} & \textbf{GitHub Project}  & \textbf{Num of\newline Retrieved Issues}  & \textbf{Num of\newline Related Issues} & \textbf{Stars}\\
\midrule
\textbf{GP1}        &  Qiskit                       & 285         & 40     &4800    &  \textbf{GP45}        &  quantum-core             & 7         & 3     &2200\\ 
\textbf{GP2}        &  Cirq                         & 212         & 35     &3900    &  \textbf{GP46}        &  qupulse          & 21          & 3     &53\\ 
\textbf{GP3}        &  deepchem                     & 148         & 23     &4600    &  \textbf{GP47}        &  Strawberry Fields         & 7          & 3      &735\\ 
\textbf{GP4}        &  ARTIQ                        & 206         & 20     &372     &  \textbf{GP48}        &  azure-quantum-python           & 6          & 2      &119\\ 
\textbf{GP5}        &  qmcpack                      & 109         & 17     &245     &  \textbf{GP49}       &  deepqmc              & 3          & 2      &300\\ 
\textbf{GP6}        &  PennyLane                    & 88          & 14      &1900     &  \textbf{GP50}       &  DFTK.jl     & 30          & 2      &356\\ 
\textbf{GP7}        &  Qiskit Metal                 & 122          & 13      &250     &  \textbf{GP51}       &  ecosystem            & 12          & 2      &70\\ 
\textbf{GP8}        &  qsharp                       & 59          & 11      &366     &  \textbf{GP52}       &  iqsharp            & 8          & 2      &126\\ 
\textbf{GP9}        &  liboqs                       & 63          & 8      &1400    &  \textbf{GP53}       &  qBraid             & 8          & 2      &62\\ 
\textbf{GP10}       &  QCoDeS                       & 24          & 8      &266     &  \textbf{GP54}       &  Qibo             & 20          & 2      &271\\ 
\textbf{GP11}       &  PyQuil                       & 31          & 7      &1400     &  \textbf{GP55}       &  Qiskit Dynamics  & 4          & 2      &96\\ 
\textbf{GP12}       &  Qiskit Nature                & 44          & 7      &239    &  \textbf{GP56}       &  QRL       & 6          & 2      &396\\ 
\textbf{GP13}       &  netket                       & 39          & 5      &450     &  \textbf{GP57}       &  QuEST& 23          & 2     &382\\ 
\textbf{GP14}       &  psi4                         & 32          & 5      &839      &  \textbf{GP58}       &  quisp & 33         & 2      &83\\ 
\textbf{GP15}       &  qsharp-language              & 24          & 5      &233     &  \textbf{GP59}       &  tensorcircuit     &10      &   2         & 250\\ 
\textbf{GP16}       &  QuTiP                        & 30           & 5      &1500      &  \textbf{GP60}       &  triqs      & 17           & 2      & 134\\ 
\textbf{GP17}       &  RMG-Py                       & 34           & 5      &329    &  \textbf{GP61}       &  Amazon Braket Python SDK               & 2           & 1      &254\\ 
\textbf{GP18}       &  Covalent                     & 48           & 4      &500     &  \textbf{GP62}       &  Catalyst    & 8           & 1      &101\\ 
\textbf{GP19}       &  cp2k                         & 49           & 4      &661     &  \textbf{GP63}       &  dftbplus & 12       & 1      &313\\ 
\textbf{GP20}       &  cuda-quantum                 & 31           & 4      &222     &  \textbf{GP64}       &  dwave-cloud-client & 1      & 1      &54\\ 
\textbf{GP21}       &  Microsoft Quantum            & 5           & 4      &3900     &  \textbf{GP65}       &  dwave-hybrid              & 1           & 1      &76\\ 
\textbf{GP22}       &  Mitiq                        & 32           & 4      &286     &  \textbf{GP66}       &  MQT DDSIM          & 1           & 1      &124\\ 
\textbf{GP23}       &  MQT QMAP                     & 7           & 4      &88      &  \textbf{GP67}       &  pennylane-qiskit      & 2           & 1      &175\\ 
\textbf{GP24}       &  QCFractal                    & 10           & 4      &143      &  \textbf{GP68}      &  ProjectQ                  & 7       & 1   & 873\\
\textbf{GP25}       &  Qrack                        & 29          & 4      &154     &  \textbf{GP69}       &  Pulser  & 14          & 1      & 159\\ 
\textbf{GP26}       &  qsharp-runtime               & 21          & 4      &280    &  \textbf{GP70}       &  QCEngine       & 2          & 1      &141\\ 
\textbf{GP27}       &  Qualtran                     & 25          & 4      &129     &  \textbf{GP71}       &  qflex & 3          & 1      &97\\ 
\textbf{GP28}       &  QuantumLibraries             & 40          & 4      &526      &  \textbf{GP72}       &  qiskit-aqua & 19         & 1      &564\\ 
\textbf{GP29}       &  Stim                         & 14          & 4      &303     &  \textbf{GP73}       &  qiskit-ibmq-provider           & 13           & 1      &228\\ 
\textbf{GP30}       &  TensorFlow Quantum           & 125         & 4      &3800      &  \textbf{GP74}       &  qiskit-ibm-runtime      & 21           & 1      &95\\ 
\textbf{GP31}       &  tket                         & 31           & 4      &197    &  \textbf{GP75}       &  qiskit-ignis              & 18           & 1      &167\\ 
\textbf{GP32}       &  circuit-knitting-toolbox     & 11           & 3      &51     &  \textbf{GP76}       &  qiskit-optimization    & 5           & 1      &215\\ 
\textbf{GP33}       &  dimod                        & 14           & 3      &120     &  \textbf{GP77}       &  qml  & 13       & 1      &509\\ 
\textbf{GP34}       &  OpenQASM                     & 18           & 3      &1000     &  \textbf{GP78}       &  qpic    & 2       & 1      &136\\ 
\textbf{GP35}       &  OpenQL                       & 15           & 3      &82     &  \textbf{GP79}       &  qsim              & 8           & 1      &361\\ 
\textbf{GP36}       &  OpenSSL                      & 16           & 3      &172     &  \textbf{GP80}       &  quacc          & 7           & 1      &155\\ 
\textbf{GP37}       &  OpenFermion                  & 17           & 3      &1500      &  \textbf{GP81}       &  QuantumKatas       & 20           & 1      &4500\\ 
\textbf{GP38}       &  PQClean                      & 16           & 3      &423      &  \textbf{GP82}      &   QUICK                 & 2       & 1   & 447\\
\textbf{GP39}       &  pyGSTi                       & 31           & 3      &131     &  \textbf{GP83}       &  Quilc          & 29           & 1      &447\\ 
\textbf{GP40}       &  PyRPL                        & 34           & 3      &114      &  \textbf{GP84}       &  SIRIUS      & 2           & 1      &103\\ 
\textbf{GP41}       &  PySCF                        & 52           & 3      &1200      &  \textbf{GP85}      &  Tangelo     & 3       & 1   & 98\\
\textbf{GP42}       &  QICK                         & 17           & 3      &175      &  \textbf{GP86}       &  toqito      & 12           & 1      &129\\ 
\textbf{GP43}       &  qiskit-metapackage           & 29           & 3      &3100      &  \textbf{GP87}      &   Yao.jl       & 14       & 1   & 907\\
\textbf{GP44}       &  qsharp-compiler              & 38           & 3      &683      &       &                    &      &    & \\
\bottomrule
\textbf{}       &                & \textbf{Total}              &        &       &        &                    & 2781     &   385 & \\

\bottomrule
\end{tabular}
\end{table}

\textbf{2) Data Collection from Stack Exchange Sites} 

To identify relevant posts in the selected Stack Exchange sites, we first conducted a pilot search to construct the search terms. The search terms were then applied to all parts of the posts, including the title, the question body, and the answers. Before we decided on the most suitable terms for capturing posts relevant to architecture of quantum software systems, we first performed a pilot search with several terms, namely ``\textit{quantum architect}'' (i.e., ``\textit{quantum architect}'', ``\textit{quantum architecture}'', ``\textit{quantum architectural}'', and ``\textit{quantum architecting}''), ``\textit{quantum design}'', and ``\textit{quantum pattern}'' within all three Stack Exchange sites. We found posts that include the keywords (``\textit{quantum architect}'', ``\textit{quantum design}'', and ``\textit{quantum pattern}'') using the original search terms. Second, Stack Exchange supports the use of wildcard (*) searches to broaden the search results. We defined ``\textit{quantum architect*}'', ``\textit{quantum design*}'', and ``\textit{quantum pattern*}'' as the initial search terms related to architecture decisions in quantum software. Table \ref{stackExchageName} shows the search terms that were utilized, and the number of posts that were obtained from all three sites. We retrieved a total of 1,494 posts. Note that a post may contain multiple sets of keywords, which means that there could be duplicate items in the set of URLs of these posts. After removing the duplicated posts, we finally got 1,424 unique posts. More specifically, 497 from Stack Overflow, 873 from Quantum Computing, and 54 from Computer Science Stack Exchange.

{\renewcommand{\arraystretch}{1}
\begin{table} [h!]
\small
\centering    
\caption{Stack Exchange Name, Search Term, Retrieved Posts, and Related Posts } \label{stackExchageName}
\begin{tabular}{p{1cm}p{3.6cm}p{3.6cm}p{2.5cm}p{2.5cm}}
\toprule
\textbf{\#}& \textbf{Stack Exchange Name}       & \textbf{Search Terms }             & \textbf{Number of Retrieved Posts}        & \textbf{Number of Related Posts}\\
\midrule
\textbf{SEN1}   & Stack Overflow                & ``\textit{quantum architect*}''                   & 148                                       & 12 \\
\textbf{SEN2}   & Stack Overflow                & ``\textit{quantum design*}''                      & 260                                       & 20 \\
\textbf{SEN3}   & Stack Overflow                & ``\textit{quantum pattern*}''                     & 115                                       & 1 \\
\textbf{SEN3}   & Quantum Computing             & ``\textit{quantum architect*}''                   & 356                                       & 10 \\
\textbf{SEN4}   & Quantum Computing             & ``\textit{quantum design*}''                      & 455                                       & 13 \\
\textbf{SEN4}   & Quantum Computing             & ``\textit{quantum pattern*}''                     & 92                                        & 10 \\
\textbf{SEN5}   & Computer Science              & ``\textit{quantum architect*}''                   & 21                                        & 2 \\
\textbf{SEN6}   & Computer Science              & ``\textit{quantum design*}''                      & 39                                        & 1 \\ 
\textbf{SEN6}   & Computer Science              & ``\textit{quantum pattern*}''                     & 8                                         & 1 \\  
\bottomrule 
\textbf{Total}  &                               &                                                   & 1494                                      & 70\\
\bottomrule
\end{tabular}
\end{table}}

\subsection{Data filtration} \label{datafiltering}
In our analysis of GitHub and Sack Exchange sites, we found many issues and posts containing the terms ``\textit{architecture}'', ``\textit{design}'' and ``\textit{pattern}'' not only in the context of software architecture and design but also other contexts, including hardware (e.g., CPU architecture\footnote{\url{https://github.com/m-labs/artiq/issues/1757}}, GPU architecture\footnote{\url{https://github.com/Qiskit/qiskit/issues/325}}, hardware design\footnote{\url{https://github.com/qiskit-community/qiskit-metal/issues/814}}, and hardware architecture\footnote{\url{https://quantumcomputing.stackexchange.com/questions/16221}}), when describing their concerns in the  GitHub issues and Stack Exchange posts. We also found that both Stack Overflow and Computer Science Stack Exchange featured numerous posts with the term ``\textit{quantum}'' not only in the context of a quantum software system but also in other contexts (e.g., Firefox Quantum\footnote{\url{https://stackoverflow.com/questions/36007119}}, quantum leap\footnote{\url{https://cs.stackexchange.com/questions/135237}}). Therefore, we need to filter the retrieved 3,185 open and closed issues from 192 quantum software systems and 1,424 posts (497 from Stack Overflow, 873 from Quantum Computing, and 54 from Computer Science Stack Exchange) and exclude those posts that are not related to software architecture decisions in quantum software systems. To do so, we performed context analysis and applied our defined inclusion and exclusion criteria (see Table \ref{mainInclusionExclusion}) to accurately filter out irrelevant issues and posts.

{\renewcommand{\arraystretch}{1}
\begin{table} [h!]
\small
\centering    
\captionsetup{font=scriptsize}
\caption{Inclusion and exclusion criteria to manually identify Stack Exchange posts and GitHub issues related to software architecture decisions in quantum software systems. }
\label{mainInclusionExclusion}
\begin{tabular}{p{15cm}}
\toprule
\textbf{Inclusion criterion}\\
\midrule
\textbf{I1.} We include a post or issue if it is related to both software architecture decisions \cite{bass2012software}\cite{van2016decision} and quantum software systems \cite{khan2023software}. For instance, if the content covers quantum software architecture patterns, design principles, structure, components, interactions, and frameworks in the development of quantum software systems, it will be included.\\	
\midrule
\textbf {Exclusion criteria}\\ 
\midrule
\textbf{E1.} We exclude a post or issue if it is only related to quantum software systems but not related to software architecture decisions. For instance, if the content covers general principles of quantum computing, basic usage of quantum programming languages, quantum hardware, or quantum physics but does not provide specific architecture decisions.\\
\textbf{E2.} We exclude a post or issue if it is only related to software architecture decisions but not related to quantum software systems. For instance, if the content covers classical software architecture, design patterns, or development methodologies that do not involve quantum computing or quantum-specific components. \\
\bottomrule 
\end{tabular}
\end{table}}

Before the formal filtering (manual inspection), we conducted a pilot filtering adhering to a set of inclusion and exclusion criteria (see Table \ref{mainInclusionExclusion}) to identify GitHub issues and Stack Exchange posts related to architecture decisions in quantum software systems. Specifically, the pilot filtering process is composed of the following steps: (1) The first author randomly selected  10 issues from all the issues containing the keywords retrieved from the selected GitHub projects and 10 posts from the search results of Stack Exchange search terms; (2) The first and third authors manually checked independently whether the issues and posts should be included by following our defined criteria (see Table \ref{mainInclusionExclusion}); (3) Data labeled by the two authors were compared, and the level of agreement between the two authors was calculated using the Cohen's Kappa coefficient \cite{cohen1960coefficient}; (4) For any posts and issues that the two authors disagreed with, the two authors discussed them till an agreement was reached. The Cohen's Kappa coefficient before the discussion and resolution of the disagreements for the pilot data filtering was 0.80 for GitHub issues and 0.76 for Stack Exchange posts, both higher than 0.7. After resolving those disagreements through discussions, the two authors reached a 100\% agreement, indicating a high degree of consistency between the two authors.

After the pilot data filtering, the first author carried on the formal data filtering with all the issues retrieved from GitHub projects and all the posts retrieved from Stack Exchange sites based on the inclusion and exclusion criteria. As it was easy to distinguish whether the posts and issues matched the criteria, the first author completed this step alone. There were only 9 issues and posts that the first author was unsure about, which were then discussed with the third author. After manually filtering all the candidates, we finally selected 385 related open and closed issues from 87 quantum software systems that related to architecture decisions in quantum software systems. These projects and their corresponding retrieved and related issues are detailed in Table \ref{gitHubProject}. Similarly, we retrieved 1,494 Stack Exchange posts and identified 70 Stack Exchange posts that discussed architecture decisions in quantum software systems. This includes 33 posts from Stack Overflow, 33 posts from Quantum Computing, and 4 posts from Computer Science sites, as detailed in Table \ref{stackExchageName}.

\subsection{Data extraction and analysis} \label{dataanalysis}

\textbf{1) Data Extraction} 

We conducted the data extraction process by identifying the relevant information to be extracted from the 385 GitHub issues and 70 Stack Exchange to answer our defined RQs (see Table \ref{table:researchQuestions}). Table \ref{dataExtraction} outlines the data items extracted from the related posts and issues with the RQs addressed using the data items. Following this, we conducted both pilot and formal data extraction.

Before the formal data extraction, the first author conducted a pilot data extraction with the second author. The first author randomly selected 10 GitHub issues and 10 Stack Exchange posts. The first two authors independently extracted relevant information (i.e., Decision description, Decision type, Application domain, Quality attribute considered, and Limitation and challenge) as detailed in Table \ref{dataExtraction}. In the event that disagreements arose, a third author was invited to discuss with the two authors and reached a 100\% agreement. After the pilot data extraction, the first author extracted the data from the filtered posts and issues independently. To mitigate personal bias, the findings from this extraction were reviewed and validated by three additional authors of this study. During this process, if any disagreements appeared, we held a meeting and followed the negotiated agreement approach \cite{campbell2013coding} to discuss and resolve any disagreement to ensure the accuracy and enhance the reliability of the extracted data regarding the uncertain portions.

\textbf{2) Data Analysis}

To analyze our extracted data by following several existing studies (e.g., \cite{zhang2023architecture}), we employed a predefined classification \cite{7371991} and the Constant Comparison method \cite{grove1988analysis} to answer RQ1 and RQ2-RQ5 in our study. To address RQ1, we employed the predefined classifications of decision description contexts as described in \cite{7371991} and \cite{zhang2023architecture}. The extracted data (i.e., decision descriptions) was then analyzed to investigate the linguistic patterns used in expressing architecture decisions. Linguistic patterns refer to grammatical rules that enable individuals to communicate effectively in a shared language \cite{da2021linguistic}. Drawing inspiration from the study of Sorbo et al. \cite{7371991}, which categorized six linguistic patterns from development emails,  we applied similar categorizations of linguistic patterns. Additionally, linguistic patterns have been employed in similar contexts, such as by Zhang \textit{et al}. \cite{zhang2023architecture}, to classify architecture decisions in AI-based systems. By applying a similar procedure (encoding and grouping of comparable codes into broader categories), we categorized linguistic patterns relevant to expressing architecture decisions to answer RQ1. Conversely, to address RQ2-RQ5, we conducted a qualitative analysis using the Constant Comparison method \cite{grove1988analysis}, which is a common technique in Grounded Theory (GT) \cite{stol2016grounded}. GT adopts a bottom-up approach aimed at generating new theories rather than extending or verifying existing ones \cite{stol2016grounded}. The constant comparison involves an ongoing process to validate the emerging concepts and categories, which are refined and saturated until they accurately represent the data \cite{stol2016grounded}\cite{grove1988analysis}. We used constant comparison to examine the different aspects of the data, including codes and categories, to unveil differences and similarities in the data \cite{hallberg2006core}.

We carried out a pilot data analysis for each RQ before the formal data analysis. The pilot data analysis encompassed several steps, as follows: (1) The first and second authors independently labeled the content of filtered posts and issues, cross-referencing them with the corresponding data items in Table \ref{dataExtraction}; (2) The first author reviewed and verified the labeling results from the second author to ensure the accuracy of data extraction; (3) The first author aggregated all codes into higher-level concepts, subsequently converting them into distinct categories; (4) To ensure the rigor of our analysis, the other five authors (the third, fourth, fifth, sixth, and seventh authors) checked and validated the results from the pilot data analysis. The disagreements were resolved through meetings, utilizing a negotiated agreement method \cite{campbell2013coding} to enhance the reliability of the pilot data analysis outcomes. In the next step, The first author carried on with the formal data analysis and followed similar steps used during the pilot data analysis. To mitigate personal bias and ensure the rigor of our analysis, all the authors checked and validated the results from the data analysis. During this process, if any disagreements appeared, the disagreements were fixed in a meeting utilizing the negotiated agreement technique \cite{campbell2013coding} to improve the reliability of the analysis outcomes, resolving any discrepancies through collaborative discussions with the first author. A summary of the data analysis approaches is provided in Table \ref{dataExtraction}. We have made all the data labeling and analysis results available in our replication package \cite{dataset}. The subsequent paragraphs outline the specifics of the formal data analysis process:

\textit{a) For analyzing RQ1}

We adapted predefined classifications from \cite{7371991} and \cite{zhang2023architecture} to address RQ1. The first author manually analyzed the extracted data for RQ1 (e.g., decision descriptions,  see Table \ref{dataExtraction}) from the related (e.g., quantum software architecture decisions) issues and posts to answering RQ1 (see Fig. \ref{OverviewOfTheResearchProcess}). The data was then examined to investigate the linguistic patterns used in expressing architecture decisions in quantum software systems. By referring to the categories of linguistic patterns presented in the previous studies, six main categories were identified. To mitigate personal bias, the identified categories were validated by all the authors of this study. The disagreements were discussed and resolved in a meeting by using the negotiated agreement approach \cite{campbell2013coding} to enhance the reliability of the results for RQ1.

\textit{b) For analyzing RQ2-RQ5}

As abovementioned, we used constant comparison \cite{grove1988analysis} to manually analyze the extracted data (i.e., the content of the issues and posts for the RQ2 decision type, the content of the issues and posts for the RQ3 application domain, the content of the issues and posts for RQ4 quality attribute
considered, and the content of the issues and posts for RQ5 limitation and challenge) as shown in Table \ref{dataExtraction}. With these RQs, we investigated architecture decisions related to issues and posts. Specifically, regarding the categorization of the questions, the first author studied the content of each post and issue related to quantum software architecture by exploring and identifying their main purposes (e.g., decision type, application domain, quality attribute, and limitation and challenge). The first author rigorously studied the data, summarized each finding into an easily understood sentence, and assigned a code. These categories, iteratively refined, encapsulated notable architecture decision types (RQ2), the different application domains where these decisions come into play (RQ3), the quality attributes developers considered during decision-making (RQ4), and the challenges faced during architecture decision-making (RQ5). Throughout the coding and categorization process, we embraced an iterative approach to enhance the accuracy of our categories. Subsequently, all the authors reviewed and validated the results of the data analysis, including the concepts and categories. Any disagreements were effectively resolved via the means of a negotiated agreement approach \cite{campbell2013coding} in meetings to mitigate personal biases. As an output, we procured categories tied to each research question, representing the richness and complexity of the extracted data. The detailed elaboration of these categories for each RQ can be found in Section \ref{results} and in Fig. \ref{ResultFigure}.

%The first author rigorously studied the data, summarizing each finding into an easily understood sentence and assigning a code. The constant comparison method was then employed to identify semantic similarities amongst codes pertinent to all research queries. Through this continuous comparative process, we could create and refine broader categories, associated with each research question. These categories, iteratively refined, encapsulated notable architectural decision types (RQ2), the different settings where these decisions come into play (RQ3), the quality attributes developers valued during decision-making (RQ4), and the challenges faced during architectural decision-making (RQ5).Throughout the coding and categorization process, we embraced an iterative approach to enhance the accuracy of our categories. To further ensure the reliability of our analysis, all authors participated in validating the generated codes, concepts, affiliating them to broader categories. Any disagreements were effectively resolved via the means of negotiated agreement approach in meetings to mitigate personal biases.As an output, we procured categories tied to each research question, representing the richness and complexity of the extracted data. The detailed elaboration of these categories for each RQ can be found in Section \ref{results} and in Fig. \ref{ResultFigure}.

{\renewcommand{\arraystretch}{1}
\begin{table} [h!]
\small
\centering  
\caption{Data items to be extracted with their description, analysis approaches, and relevant RQs} 
\label{dataExtraction}
\begin{tabular}{p{0.5cm}p{3cm}p{5.7cm}p{3cm}p{1cm}}
\toprule
\textbf{\#} & \textbf{Data item}                   & \textbf{Description}         & \textbf{Data analysis approach}                                           & \textbf{RQs}  \\ 
\midrule
D1 & Decision description                          & \textit{Developers-based description of architecture decisions made in the development of quantum software systems} 
                                                                                  & Predefined classification \cite{7371991}                                  & RQ1           \\ \hline		
D2 & Decision type                                 & \textit{A classification of architecture decisions taken during the development of quantum software systems} 
                                                                                  & Constant comparison                                                       & RQ2           \\ \hline		
D3 & Application domain                            & \textit{Application domains in which architecture decisions are made throughout the development of quantum software systems} 
                                                                                  & Constant comparison                                                       & RQ3           \\ \hline		
D4 & Quality attribute considered                  & \textit{Considered quality attributes when developers of quantum software systems make architecture decisions} 
                                                                                  & Constant comparison                                                       & RQ4           \\ \hline	
D5 & Limitation and challenge                      & \textit{The obstacles and difficulties posed by quantum software development when making architecture decisions} 
                                                                                  &  Constant comparison                                                      & RQ5           \\ 
\bottomrule 
\end{tabular}
\end{table}}

\section {Results} \label{results}
\begin{figure}[h]
 \centering
 \includegraphics[width=1.0\linewidth]{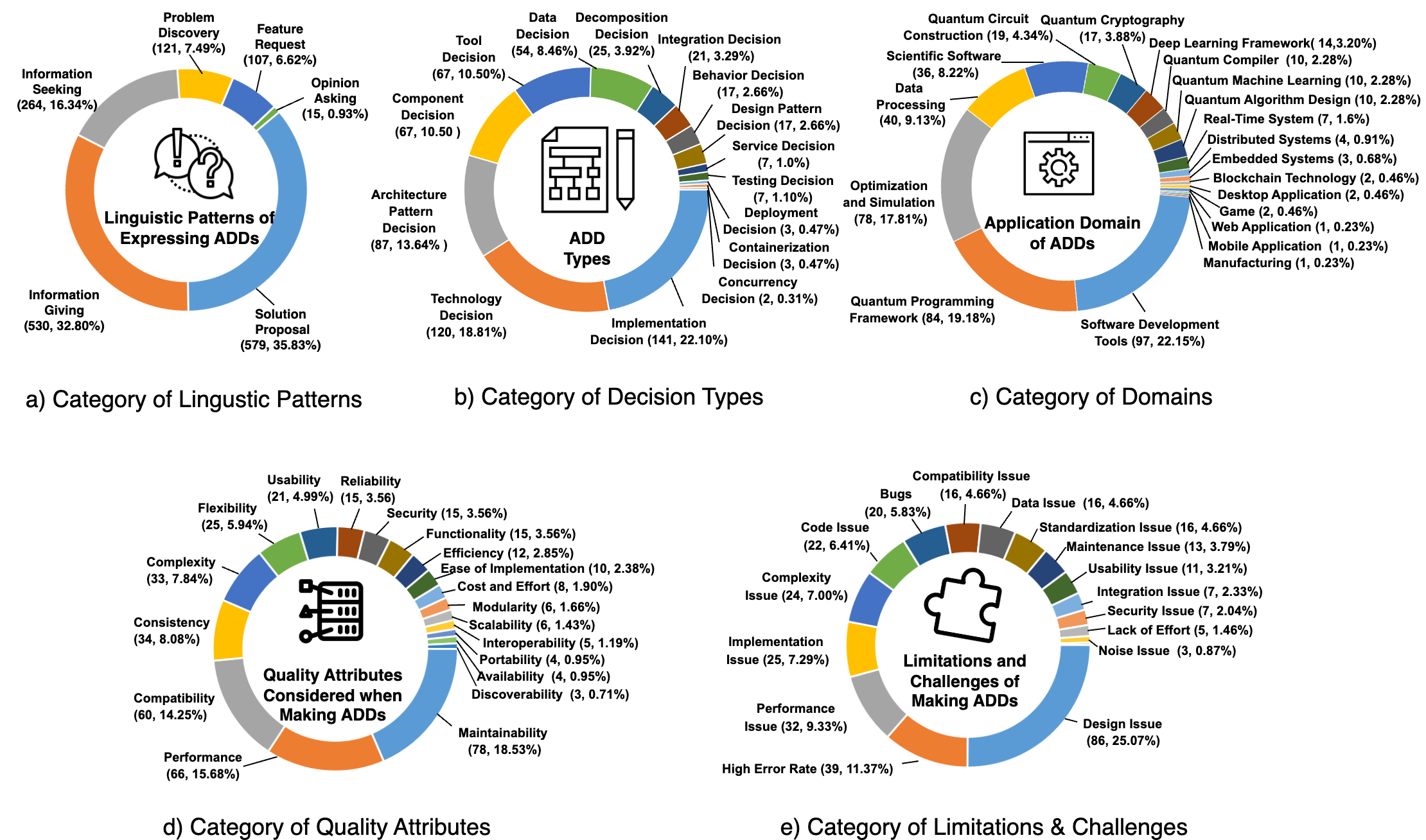}
 \caption{Linguistic Patterns, Architecture Decisions, Application Domains, Quality Attributes, and Limitations\&Challenges (results of RQ1 to RQ5) }
 \label{ResultFigure}
\end{figure}

In this section, we present the results of the five RQs as outlined in Table \ref{table:researchQuestions}. We first describe our methodology for analyzing instances drawn from GitHub issues and Stack Exchange discussions, and we then present the study results of each RQ. The results of the five RQs are visualized in Fig. \ref{ResultFigure}.

\subsection{RQ1. How do developers express architecture decisions made in quantum software development?} 
\label{resultsofRQ1}

Our approach to answering RQ1 involves multiple stages. First, we collected 1616 instances of linguistic patterns (1406 from GitHub and 210 from Stack Exchange sites) of how practitioners articulate architecture decisions during the development of quantum software systems. One GitHub issue and one Stack Exchange post may identify multiple linguistic patterns because, within a single issue or post, practitioners might articulate their architecture decisions in various ways, using different phrases, terminologies, or contexts. To classify these decision expressions according to their intended objectives, we utilized a predefined classification scheme based on the linguistic patterns proposed by Sorbo et al.~\cite{7371991}. In a similar context, Zhang \textit{et al}. \cite{zhang2023architecture} employed linguistic patterns to classify architecture decisions in the development of AI-based systems. Ultimately, we categorized the extracted architecture decision expressions into six types of linguistic patterns. Fig. \ref{ResultFigure}a presents the six linguistic patterns and Table \ref{linguisticPatterns} contains these linguistic patterns as well as their examples and respective percentages.

Our analysis revealed that \textit{Solution Proposal} and \textit{Information Giving} were the two most common linguistic patterns, accounting for 35.83\% and 32.80\% of the instances, respectively. In contrast, \textit{Opinion Asking} and \textit{Problem Discovery} received relatively less attention from practitioners when conveying architecture decisions within the context of quantum software systems. A detailed overview of each of those categories is provided below: 

{\renewcommand{\arraystretch}{1}
\begin{table} [h!]
\small
\centering  
\caption{Linguistic patterns of architecture decisions in quantum software systems}
\label{linguisticPatterns}

\begin{tabular}{p{3cm}p{8.2cm}p{1cm}p{1cm}}
\toprule
\textbf{Linguistic Pattern}     &  \textbf{Example}                  & \textbf{Count}           & \textbf{\%}         \\ \hline 
                
Solution Proposal               & \textit{One solution (to do it generically) that would involve a somewhat major architectural change is to implement a generic interface to asymmetric encryption then implement KEX in terms of that for the algorithms it applies to.} (liboqs issue \#50)                                                                                                                                                         &  579                     & 35.83\%               \\ \hline
Information Giving              & \textit{Currently the hybrid Python/C++ architecture of Psi4 is in an odd spot where Psi4 itself is a C++ program that calls an input file as a Python executable. This circular process will be removed and Psi4 itself will become a Python library capable of being imported just like any other module.}  (psi4 issue \#468)                  
                                                                     &  530                    & 32.80\%              \\ \hline
Information Seeking             & \textit{After we have finished the decomposition and local Covalent has a microservice architecture, what should happen when a user does covalent start?}  (Covalent issue \#186)                       
                                                                     &  264                    & 16.34\%              \\ \hline
Problem Discovery               & \textit{These two software packages are very problematic: Conda is very poorly architected/designed.} (ARTIQ issue \#1471)      
                                                                     &  121                    & 7.49\%              \\ \hline    
Feature Request                 & \textit{Architectures supported in liboqs master branch should be updated} (liboqs issue \#751)                                          
                                                                     &  107                     & 6.62\%              \\ \hline
Opinion Asking                  & \textit{Please feel free to suggest changes or alternate designs.} (deepchem issue \#2059)      
                                                                     &  15                      & 0.93\%               \\ 
\bottomrule 
\end{tabular}%
\end{table}}

%Each linguistic pattern provides a unique concept through which practitioners express architecture decisions in the quantum software domain. A detailed exploration of these patterns provides valuable insights into how developers communicate and approach architecture decisions. Therefore, we provide more details below.

\textbf{(1) Solution Proposal}: Our analysis shows that \textit{Solution Proposal} is the most frequently used linguistic pattern. Those solutions are offered as pertinent information or assistance to fellow developers. This category often encompasses suggestions, recommendations, and best practices tailored to support and guide peers within the community. In our analysis, developers mentioned \textit{Solution Proposal} in about 35.83\% of the linguistic patterns we collected.

\textbf{(2) Information Giving}: This category relates to developers sharing insights, facts, or knowledge about architecture decisions. Developers might elucidate the nuances of a quantum algorithm, describe the interaction between classical and quantum components, or explain design choices made to optimize quantum coherence. This category embodies the proactive dissemination of architectural knowledge, ensuring that team members are aligned and informed about the architecture decisions made. In our analysis, developers mentioned \textit{Information Giving} in about 32.80\% of the linguistic patterns we collected.

\textbf{(3) Information Seeking}: \textit{Information Seeking} revolves around developers looking to enhance product quality or gain clarity on specific functionalities within quantum software systems. This might manifest as questions, requests for clarification, or discussions aiming to understand architecture decisions better. In our analysis, developers mentioned \textit{Information Seeking} in about 16.34\% of the linguistic patterns we collected.

\textbf{(4) Problem Discovery}: In \textit{Problem Discovery} linguistic pattern, developers identify, share, and discuss new problems, bugs, or issues they encounter in quantum software systems. It highlights the initial stages of problem-solving, where issues are first recognized and made known to the community. In our analysis, developers mentioned \textit{Problem Discovery} in about 7.49\% of the linguistic patterns we collected.

\textbf{(5) Feature Request}: When making decisions about the architecture design of quantum software systems, developers may ask for new features or enhancements to already existing libraries and software tools. These may also include recommendations to improve the functionality, usability, or flexibility of the systems. In our analysis, developers mentioned \textit{Feature Request} in about 6.62\% of the linguistic patterns we collected.

\textbf{(6) Opinion Asking}: \textit{Opinion Asking} involves developers seeking the views, thoughts, or sentiments of peers concerning architecture decisions in quantum software systems. It may range from inquiring about the practicality of a proposed solution to gauging how the community feels about specific architectural decisions. In our analysis, developers mentioned \textit{Opinion Asking} in about 0.93\% of the linguistic patterns we collected.

\begin{tcolorbox}[colback=gray!5!white,colframe=gray!75!black,title=Key Findings of RQ1]
\textbf{Finding 1}:
Developers represent architecture decisions with six categories of linguistic patterns, among which \textit{Solution Proposal} (35.83\%) and \textit{Information Giving} (32.80\%) are the two most frequently employed categories in quantum software development (see the examples in Table \ref{linguisticPatterns}). 
\end{tcolorbox}

\subsection{RQ2. What types of architecture decisions are made in quantum software development?} 
\label{resultsofRQ2}

To answer RQ2, we conducted a comprehensive review of the content pertinent to architecture decisions in the context of quantum software systems, as gleaned from the accumulated 638 instances of architecture decisions (553 from GitHub and 85 from Stack Exchange sites). Note that a single post or issue often contains multiple architecture decisions because developers often explore various aspects of a problem, seeking information for comprehensive solutions or gathering diverse opinions. Additionally, the same architecture decision can be discussed multiple times within a post or issue due to its significance or as a common challenge faced by developers, leading to repetitive exploration for clarity or confirmation. Therefore, if a post or issue mentions multiple architecture decisions, each decision is counted. However, repeated mentions of the same architecture decision within a single post or issue were only counted once. We used the Constant Comparison method \cite{grove1988analysis} to categorize the architecture decisions pertinent to the development of quantum software systems. Fig. \ref{ResultFigure}b presents 15 types of architecture decisions and Table \ref{typeOfArchitectureDecision} provides an overview of these types of architecture decisions made in quantum software development, along with their respective quantities and percentages.

As shown in Table \ref{typeOfArchitectureDecision}, the two common types of decisions are \textit{Implementation Decision} (22.10\%) and \textit{Technology Decision} (18.81\%). The remaining categories of architecture decisions account for a small percentage of the architecture decisions (less than 15\%). We provide a detailed description of each of those categories below: 

{\renewcommand{\arraystretch}{1}
\begin{table} [h!]
\small
\centering  
\caption{Types of architecture decisions in quantum software development and their counts \& percentages}
\label{typeOfArchitectureDecision}
\begin{tabular}{p{3cm}p{8.2cm}p{1cm}p{1cm}}
\toprule
\textbf{Decision Type}               &  \textbf{Example}                       & \textbf{Count}                         & \textbf{\%} \\ \hline 
Implementation Decision              &  \textit{We designed and implemented an architectural framework that aims to standardize the interaction of an existing infrastructure with IBM Quantum, leveraging the most used open source technologies.} (ecosystem issue \#266)                  
                                                                               &  141                                 & 22.10\%   \\ \hline                
Technology Decision                  & \textit{Also, since the technology of partial measurement/non-demolition measurement is under development (I believe in some quantum architectures), may I expect to use this SDK to change when the technology is mature?} (amazon-braket-sdk-python issue \#158)          
                                                                               &  120                                  & 18.81\%    \\ \hline
Architecture Pattern Decision        & \textit{If we are willing to provide extensibility (which sounds reasonable in a pluggable ecosystem), we should shift to a more suitable pattern such as the observer or pub-sub pattern (...).Again, the pub/sub or obeserver patterns are examples of decoupled architectures.} (Qiskit \#825)      
                                                                               &  87                                   & 13.64\%      \\ \hline 
Component Decision                   &  \textit{If we want to be strict about the architecture, the log forwarding component should be a controller.} (artiq issue \#691)                 
                                                                               &  67                                   & 10.50\%    \\ \hline   
Tool Decision                        & \textit{Unfortunately this is something coming from Amber build system. It compiles cuda source files for all possible architectures that a given toolkit supports.} (QUICK issue \#218)                                                                         
                                                                               & 67                                    & 10.50\%      \\ \hline       
Data Decision                        & \textit{This seems like part of the discussion about long term architecture, harmonizing dc.data with tf.data and TensorGraph layers with Keras layers.} (deepchem issue \#1036)      
                                                                               & 54                                    & 8.46\%      \\ \hline 
Decomposition Decision               &  \textit{After we have finished the decomposition and local Covalent has a microservice architecture, what should happen when a user does covalent start?} (covalent issue \#186) 
                                                                               & 25                                   & 3.92\%       \\ \hline 
Integration Decision                 &  \textit{I will try integrating quantum layers into existing CNN architectures} (QC post \#33215)      
                                                                               & 21                                   & 3.29\%      \\ \hline  
Behavior Decision                    &  \textit{I would propose not keeping the old behavior, simply to make the refactor easier and avoid code complexity (as well as design decisions).} (PennyLane issue \#226)      
                                                                               & 17                                 & 2.66\%       \\ \hline 
Design Pattern Decision              & \textit{Builder Pattern for defining control problems(...).The optimize\_pulse API takes in a lot of optional arguments. Wouldn't it make more sense to use a builder pattern to define a control problem in a ControlProblem class?} (qutip issue \#692)     
                                                                               &   17                                   & 2.66\%       \\ \hline 
Service Decision                     & \textit{The sync method in covalent/\_results\_manager/results\_manager.py needs to be moved and updated to be compatible with the new covalent services architecture.} (Covalent issue  \#370)                                                                                                                                                                              &   7                                   & 1.10\%       \\ \hline 
Testing Decision                    & \textit{The assert for the third derivatives in test\_einset.cpp fails due to what looks like some finite precision issue, but the issue can't be reproduced on other architectures. Unit test for third\-derivative components disabled until this issue is identified.} (QMCPACK issue  \#1218)                                                                                                                                                                             &   7                                   & 1.10\%      \\ \hline 
Deployment Decision                  &  \textit{QPIC is designed to be compatible from Python 2 up to the latest Python 3 (...). Is already fixed in this repo but not deployed to PyPI. Could you update the package in PyPI?} (qpic issue \#12)      
                                                                               &  3                                     & 0.47\%       \\ \hline 
Containerization Decision            &  \textit{I agree that there is a problem (smell) with the design. We are treating the amount of work as related to the size of the container when with padding this is not the case.} (qmcpack issue \#4290)     
                                                                               &  3                                     & 0.47\%       \\ \hline 
Concurrency Decision &  \textit{The architecture implemented at the moment is showing some issues that are non trivial to fix  (...).and then carefully design how we want to do concurrency and parallelism.} (Qcodes issue \#272)                                                                                                    &   2                                                     & 0.31\%                                          \\                                                                              
\bottomrule 
\end{tabular}%
\end{table}}
%While our table presents a quick quantitative glance at decision distribution, diving deeper into each decision provides essential context, bridging the gap between theory and practical implications. Such in-depth descriptions give developers insights into the intricate considerations behind each choice, ensuring a holistic grasp of the complexities in quantum software design. Therefore, we provide more details of each architecture decision below.

\textbf{(1) Implementation Decision:} Developers need to make informed decisions about implementing quantum algorithms and operations within their systems. These decisions often involve intricate technical considerations, such as selecting suitable programming languages and quantum computing platforms and optimizing code for quantum hardware. \textit{Implementation Decisions} are found in 22.10\% of the architecture decisions we collected.

\textbf{(2) Technology Decision}: Developers frequently find themselves at a crossroads, needing to decide which particular quantum technologies should be used due to the rapidly developing aspect of quantum technologies. These decisions have far-reaching implications for the overall system's efficiency and functionality, as selecting the right technology can significantly impact the system's performance. We found that \textit{Technology Decisions} represent 18.81\% of the architecture decisions we collected.

\textbf{(3) Architecture Pattern Decision}: \textit{Architecture Pattern Decision} refers to the high-level organizational structure and modular design choices developers adopt when designing quantum software or integrating quantum components within larger systems. We identified the most common architecture patterns in quantum software systems as qubit gate patterns, layer patterns, composite patterns, decorator patterns, pipe and filter patterns, and transformer patterns. These could include modular quantum algorithm design, pipeline structures for quantum operations, or layered approaches separating quantum and classical components. Just as classical systems might leverage patterns like layers or service orientation, quantum systems require patterns optimized for quantum mechanics. These patterns shape the workflow, scalability, and maintainability of quantum systems. In a review of architecture decisions made in quantum software systems, \textit{Architecture Pattern Decisions} were discovered in 13.64\% of the architecture decisions we collected.

\textbf{(4) Component Decision}: Quantum software systems, like other software systems, benefit from modular and well-structured designs. Developers often need to decide how to organize and design system components, considering factors like quantum modules, quantum layers, and the integration of quantum and classical components. \textit{Component Decision} is integral to achieving scalability, maintainability, and overall system reliability. \textit{Component Decisions} were identified in 10.50\% of our collected architecture decisions.

\textbf{(5) Tool Decision}: Quantum systems utilize a myriad of tools tailored for quantum-specific tasks. In the process of developing quantum software, developers commonly have to make an essential decision: choosing the right toolset or frameworks. This decision deeply influences the system's overall performance, fidelity in simulation, ease of debugging, and efficient resource orchestration. \textit{Tool Decisions} were found in 10.50\% of the architecture decisions we collected. 

\textbf{(6) Data Decision}: \textit{Data Decision} refers to how quantum data structures and quantum bits (qubits) are managed, accessed, and manipulated. Given the inherent differences between classical and quantum data, determining the right strategies for quantum data storage, retrieval, and processing is essential. This type of decision encompasses selecting appropriate quantum algorithms, choosing between quantum states, and handling superposition and entanglement. \textit{Data Decisions} were discovered at 8.46\% of the architecture decisions we collected.

\textbf{(7) Decomposition Decision}: \textit{Decomposition Decision} pertains to the choices made when breaking down complex quantum algorithms or processes into simpler and more manageable components. Quantum algorithms, given their intricate nature, often need to be decomposed into a series of quantum gates or subroutines. The decision on how to decompose not only affects the efficiency of the algorithm but also its compatibility with specific quantum hardware. Such decisions may also involve deciding the granularity of modules or components, ensuring that the architecture remains modular, scalable, and maintainable. \textit{Decomposition Decisions} were identified in 3.92\% of the architecture decisions we collected. 

\textbf{(8) Integration Decision}: \textit{Integration Decision} refers to decisions made concerning how various software components, modules, or subsystems communicate and work together within quantum software systems. These decisions can involve the selection of quantum libraries, ensuring compatibility between classical and quantum sub-systems, or determining how quantum algorithms interface with other software functionalities. \textit{Integration Decisions} were discovered at 3.29\% of the architecture decisions we collected. 

\textbf{(9) Behavior Decision}: \textit{Behavior Decision} revolves around determining how quantum components or modules will interact and behave over time. Given that quantum software systems are inherently probabilistic and are susceptible to decoherence and external interference, \textit{Behavior Decisions} often involve trade-offs between performance, accuracy, and fault tolerance. \textit{Behavior Decisions} were discovered at 2.66\% of the architecture decisions we collected.

\textbf{(10) Design Pattern Decision}: \textit{Design Pattern Decision} in quantum software systems involves strategically selecting and applying design patterns to address common architecture challenges specific to quantum computing. These patterns help structure the software systems in a maintainable, scalable, and efficient way, addressing challenges such as state management, circuit construction, and operation optimization. By understanding and applying these well-known design patterns (e.g., visitor, builder, and recursive patterns) developers can ensure that their quantum software systems are built on robust architectural foundations. \textit{Design Pattern Decisions} were discovered at 2.66\% of the architecture decisions we collected.

\textbf{(11) Service Decision}: \textit{Service Decision} pertains to the orchestration of quantum computational tasks within software frameworks. This involves decisions related to offering quantum functionalities as modular services or APIs. Considerations of \textit{Service Decision}  include the encapsulation of specific quantum algorithms, the design of interfaces for quantum service interaction, and the efficient management of quantum data structures. The focus is on the structural organization of quantum services within the software, ensuring modularity, scalability, and robustness while maintaining the peculiarities of quantum computation. \textit{Service Decisions} were identified in 1.10\% of the architecture decisions we collected.

\textbf{(12) Testing Decision}: \textit{Testing Decision} in quantum software architecture refers to the process of selecting and designing appropriate testing strategies and practices to verify and validate the functionality, performance, and reliability of quantum software systems. This involves choosing the types of tests (unit tests, integration tests, etc.), identifying the scope of each test, and determining how to handle specific challenges related to quantum computing, such as finite precision issues, architecture-specific behaviors, and the infeasibility of exhaustive testing for large qubit systems. \textit{Testing Decisions} were identified in 1.10\% of the architecture decisions we collected.

\textbf{(13) Deployment Decision}: \textit{Deployment Decision} encapsulates strategies and choices regarding how and where quantum computing applications will be executed and utilized. Further, deployment decisions extend to defining how quantum solutions are made available to end-users or other systems, ensuring that the deployment aligns with the computational, security, and performance requirements of quantum applications. \textit{Deployment Decisions} were found in 0.47\% of the architecture decisions we collected. 

\textbf{(14) Containerization Decision}: \textit{Containerization Decision} in a quantum software system involves the strategic use of containers to encapsulate and manage the software components and their dependencies. These decisions focus on ensuring the portability, scalability, and consistency of the quantum software across different environments. Containerization enables developers to package quantum applications with all necessary libraries and dependencies, ensuring that the application runs seamlessly on any platform that supports containers. \textit{Containerization Decisions} were found in 0.47\% of the architecture decisions we collected.

\textbf{(15) Concurrency Decision}: \textit{Concurrency Decision} is concerned with managing multiple quantum operations or processes within software frameworks. \textit{Concurrency Decisions} must consider the potential simultaneous execution of quantum tasks. This ensures that the quantum software efficiently utilizes available quantum resources while ensuring the integrity and fidelity of quantum computations. \textit{Concurrency Decisions} were discovered at 0.31\% of the architecture decisions we collected.

\begin{tcolorbox}[colback=gray!5!white,colframe=gray!75!black,title=Key Findings of RQ2]
\textbf{Finding 2}:
\textit{Implementation Decision} and \textit{Technology Decision} are the most common decision types when making architecture decisions in quantum software development. Specifically, \textit{Implementation Decisions} account for 22.10\%, \textit{Technology Decision} account for 18.81\% of the architecture decisions we collected (see more details with examples
in Table \ref{typeOfArchitectureDecision}).
\end{tcolorbox}

\subsection{RQ3. What are the application domains of quantum software in which architecture decisions are made?} 
\label{resultsofRQ3}

To answer RQ3, we analyzed the application domains where architecture decisions are made in the context of the development of quantum software systems. The finding of application domains is presented in Fig. \ref{ResultFigure}c and summarized in Table~\ref{domainArchitectureDecisions}, which contains 438 instances of application domains in total (373 from GitHub issues and 65 from Stack Exchange sites). During the data analysis process, we identified one application domain from each post and issue because our focus was on the primary application domain in which an architecture decision was made in quantum software systems, and most of the time the same domain was mentioned multiple times within a post or an issue, and we rarely found one post or issue that mentioned multiple domains where architecture decisions were made. 

We identified 20 application domains, in which \textit{Software Development Tools} are dominant, accounting for 22.15\% of the application domains. \textit{Quantum Programming Framework} represents 19.18\% of the application domains. The remaining application domains account for less than 18\% of the application domains. A detailed description of the top 9 application domains is provided below.

% \textit{Scientific Software}, in addition to \textit{Optimization and Simulation}, \textit{Data Processing}, and \textit{Quantum Cryptography}, are among 13\% and 7\%. Additionally, other application domains account for less the 5\%. \textit{Deep Learning Framework}, \textit{Quantum Compiler}, and \textit{Real-Time System} are between 4\% and 3\%. \textit{Distributed System}, \textit{Embedded System}, \textit{Blockchain Technology}, and \textit{Desktop Application}, account for around 1\% of the overall sector. The remaining application domains, \textit{Web Application}, \textit{Mobile Application}, and \textit{Game} are each mentioned only once, accounting for less than 1\%.

{\renewcommand{\arraystretch}{1}
\begin{table} [h!]
\small
\centering  
\caption{The domains of quantum software systems in which architecture decisions are made}
\label{domainArchitectureDecisions}
\begin{tabular}{p{5.2cm}p{0.9cm}p{0.9cm}|p{4.4cm}p{0.9cm}p{0.9cm}}
\toprule
\textbf{Domain} &  \textbf{Count} & \textbf{\%} & \textbf{Domain} & \textbf{Count} & \textbf{\%} \\ \hline                 
Software Development Tools                  & 97    &   22.15\%      & Quantum Algorithm Design     & 10    & 2.28\%    \\ \hline
Quantum Programming Framework               & 84    &   19.18\%       & Real-Time System                                & 7     & 1.6\%     \\ \hline
Optimization and Simulation                 & 78    &   17.81\%      & Distributed Systems                             & 4     & 0.91\%    \\ \hline
Data Processing                             & 40    &   9.13\%       & Embedded Systems                                & 3     & 0.68\%     \\ \hline
Scientific Software                         & 36    &   8.22\%       & Blockchain Technology                           & 2     & 0.46\%    \\ \hline
Quantum Circuit Construction                & 19    &   4.34\%       & Desktop Application                             & 2     & 0.46\%     \\ \hline
Quantum Cryptography                        & 17    &   3.88\%       & Game                                            & 2     & 0.46\%     \\ \hline 
Deep Learning Framework                     & 14    &   3.20\%       & Web Application                                 & 1     & 0.23\%    \\ \hline 
Quantum Compiler                            & 10    &   2.28\%       & Mobile Application                              & 1     & 0.23\%    \\ \hline
Quantum Machine Learning                    & 10    &   2.28\%       & Manufacturing                                   & 1     & 0.23\%    \\ 
\bottomrule 
\end{tabular}
\end{table}}

\textbf{(1) Software Development Tool}: Tools are pivotal for designing and building quantum software systems. These tools, ranging from Integrated Development Environments (IDEs) to debugging platforms, ensure efficient code development, testing, and deployment. In a quantum context, these tools are specialized to handle quantum-specific challenges, like qubit manipulation and entanglement. \textit{Software Development Tool} was found at 22.15\% of the application domains we collected. 

\textbf{(2) Quantum Programming Framework}: The most common domain is revealed to be the \textit{Quantum Programming Framework}. This domain holds substantial importance in quantum software systems, as it forms the foundation for the development of quantum algorithms and applications. It enables programmers to harness the power of quantum computation by providing the necessary tools, libraries, and abstractions, thus facilitating the creation of quantum software solutions. \textit{Quantum Programming Framework} was found in 19.18\% of the application domains we collected. 

\textbf{(3) Optimization and Simulation}: optimization problems are central to quantum software systems, aiming to model and improve real-world quantum scenarios. By simulating quantum environments, developers can test algorithms or explore quantum behavior, making decisions about system robustness, efficiency, and accuracy. \textit{Optimization and Simulation} instances were found at 17.81\% of the application domains we collected.

\textbf{(4) Data Processing}: \textit{Data Processing} emphasizes handling and analyzing quantum data, which can be vastly different from classical data. Architecture decisions here revolve around quantum data structures, efficient algorithms for quantum data manipulation, and ensuring data integrity. \textit{Data Processing} was identified at 9.13\% of the application domains we collected.

\textbf{(5) Scientific Software}: \textit{Scientific Software} in quantum systems deals with applications tailored for research and complex computations. These tools are designed to understand and utilize quantum phenomena, aiding scientists in areas like quantum physics, chemistry, and material science. The architectural decision here focuses on precision, scalability, and accuracy. \textit{Scientific Software} was found at 8.22\% of the application domains we collected.

\textbf{(6) Quantum Circuit Construction}: \textit{Quantum Circuit Construction} in quantum software systems involves the systematic design, creation, and optimization of quantum circuits, leveraging architectural patterns and frameworks that facilitate hardware-software co-organization. This process integrates considerations of both the logical design of quantum algorithms and the physical constraints and capabilities of the quantum hardware, ensuring efficient and reliable execution of quantum computations. \textit{Quantum Circuit Construction} was found at 4.34\% of the application domains we collected.

\textbf{(7) Quantum Cryptography}: This domain is primarily concerned with utilizing the principles of quantum mechanics to secure communication against any potential interception. It involves techniques like Quantum Key Distribution (QKD), where cryptographic keys are securely distributed, and eavesdropping can be detected. The main architectural decisions in quantum cryptography revolve around ensuring the security of data during transfer. It is about securing communication channels in a way that is theoretically immune to any computational hacks, including those using quantum computers. \textit{Quantum Cryptography} was found at 3.88\% of the application domains we collected. 

\textbf{(8) Deep Learning Framework}: \textit{Deep Learning Framework} can help facilitate the fusion of quantum computing and neural networks. Such frameworks allow quantum systems to process vast amounts of data rapidly, making decisions about quantum neural architectures, training methods, and data-feeding mechanisms. \textit{Deep Learning Framework} was found at 3.20\% of the application domains we collected. 

\textbf{(9) Quantum Compiler}: Those specialized compilers translate high-level quantum programming languages into machine-level instructions for quantum processors. Decisions in this domain concern optimization techniques, error corrections, and ensuring the efficient execution of quantum operations. \textit{Quantum Compiler} was found at 2.28\% of the application domains we collected.

\begin{tcolorbox}[colback=gray!5!white,colframe=gray!75!black,title=Key Findings of RQ3]
\textbf{Finding 3}:
Our analysis of application domains of quantum software in which architecture decisions are made classified a total of 438 instances into 20 categories. \textit{Software Development Tools} are dominant in these categories, accounting for 22.15\% of the application domains we collected (see more details with examples in Table \ref{domainArchitectureDecisions}).
\end{tcolorbox}

\subsection{RQ4. What quality attributes are considered when developers make architecture decisions in quantum software development?} 
\label{resultsofRQ4}
To answer RQ4, we extracted data concerning \textit{quality attributes}. We collected 421 instances of quality attributes (376 from GitHub Issues and 45 from Stack Exchange posts). We discovered that a single post or issue mentioned multiple quality attributes when discussing architecture decisions because quantum software systems, being inherently complex, require developers to consider several quality attributes simultaneously to ensure robust and effective architecture decisions. If we identified the same type of quality attributes within the same post or issue, we considered them only once. Subsequently, we categorized these instances into 19 types of quality attributes, which are presented in Fig. \ref{ResultFigure}d and detailed in Table \ref{qualityAttribute}. 

Among these quality attributes, \textit{Maintainability} appears as the most frequently considered by practitioners, constituting 18.53\% of the total instances. Furthermore, \textit{Performance} and \textit{Compatibility} are accounting for 15.68\% and 14.25\%. Additionally, other quality attributes fall below the 10\% mark, signifying that these quality attributes are the least considered by practitioners in their architecture decision-making of quantum software development. A detailed description of the top 7 quality attributes is provided below.

{\renewcommand{\arraystretch}{0.9}
\begin{table} [!]
\small
\centering  
\caption{The quality attributes considered when making architecture decisions in quantum software development and their counts \& percentages}
\label{qualityAttribute}
\begin{tabular}{p{3cm}p{8.2cm}p{1cm}p{1cm}}
\toprule
                    \textbf{Quality Attribute}                         &  \textbf{Example}                       & \textbf{Count}                        & \textbf{\%}      \\ \hline 
Maintainability          &  \textit{It would mean a more confused internal architecture, which would hurt us in maintainability in the long term.} (Qiskit issue \#1852)                                            
                                                                                                                 &  78                                   & 18.53\%           \\ \hline 
Performance              &  \textit{First the current design is intrinsically bad performance for hot loops because we have to look up the state (even if it doesn't change).} (DFTK.jl issue \#313)                
                                                                                                                 &  66                                   & 15.68\%           \\ \hline
Compatibility            &  \textit{The ``my foundational program can do many things, how do I box its capabilities where the QCArchive architecture views it as separate pieces''.} (QCEngine issue \#197)         
                                                                                                                 &  60                                   & 14.25\%          \\ \hline  
Consistency              &  \textit{Ensure the API is as consistent as possible (both between PyTorch based models.} (deepchem issue \#2059)                                                                        
                                                                                                                 &  34                                   & 8.08\%             \\ \hline 
Complexity               &  \textit{In order to keep Psi4 up to date with best C++11 practices, reduce complexity (...). Currently the hybrid Python/C++ architecture of Psi4 is in an odd spot where Psi4 itself is a C++ program that calls an input file as a Python executable.} (psi4 issue \#468)                                                                                                                                                                                                                                                                                                                 &  33                                  & 7.84\%             \\ \hline 
Flexibility              &  \textit{I think RMG should be designed to not require changes to the core functionality, but have various inherited classes that allow for flexibility and robustness.} (RMG-Py issue \#1136)      
                                                                                                                 &  25                                  & 5.94\%            \\ \hline 
Usability                &  \textit{D-Wave Hybrid is a (Python) framework designed for creating classical-quantum hybrid workflows with an emphasis on ease of use.} (dwave-hybrid issue \#245)                     
                                                                                                                 &  21                                  & 4.99\%             \\ \hline
Reliability              &  \textit{MSYS also has straightforward benefits for users, such as speed and reliability.} (ARTIQ issue \#1471)                                                                          
                                                                                                                 &  15                                   & 3.56\%             \\ \hline
Security                 &  \textit{I think it is a very important project for the future of our security: being able to work with these new algorithms.} (OpenSSL issue  \#295)                                    
                                                                                                                 &  15                                  & 3.56\%             \\ \hline 
Functionality                 &  \textit{The Q\# Formatter should have a design specification document, written in Markdown like a README, that describes the functionality of the tool as we want it to be and details our goals for the tool.} (qsharp\-compiler issue \#1074)                                   
                                                                                                                 &  15                                   & 3.56\%            \\ \hline 
Efficiency                &  \textit{Samplers use Pauli frames to efficiently handle noisy circuits behind the scenes, but hide this information in the Python API.} (Stim issues \#135)                                   
                                                                                                                 &  12                                   & 2.85\%             \\ \hline 
Ease of Implementation   &  \textit{I think we all agree that caching is a good idea; the question is just how easy it is to implement in the current architecture.} (ARTIQ issue \#1542)                           
                                                                                                                 &  10                                   & 2.38\%             \\ \hline 
Cost and Effort          &  \textit{Lots of money has been poured already into Conda (Continuum Analytics, NumFocus) and VS (Microsoft) with such a shoddy final result, so those tools seem like a dead-end to me.} (ARTIQ issue \#1471)      
                                                                                                                 &  8                                    & 1.90\%             \\ \hline 
Modularity               &  \textit{We have a good modularity in the compiler, specifically with respect to the distinctions above.} (cuda-quantum issue \#85)                                                      
                                                                                                                 &  6                                    & 1.66\%            \\ \hline 
Scalability              &  \textit{Using db will solve the scalability issue and will be able to handle a large number of request.} (QRL issue \#1545)                                                             
                                                                                                                 &  6                                    & 1.43\%            \\ \hline 
Interoperability         &  \textit{We also get better interoperability with the broader python ML ecosystem.} (deepchem issue \#/1036)                                                                             
                                                                                                                 &  5                                    & 1.19\%             \\  \hline  
Portability              &  \textit{dc.data is one of the hackier parts of DeepChem and has caused us a good number of scaling and portability headaches.} (deepchem issue \#1039)                                  
                                                                                                                 &  4                                    & 0.95\%             \\ \hline  

Availability             &  \textit{Availability of QML tools such as optimizers and different encoding schemes`` and ''availability of simulators and circuit visualizers.} (QC post \#33215)                      
                                                                                                                 &  4                                    & 0.95\%             \\ \hline
Discoverability          &  \textit{In addition to the general Q\# API Design Principles, the changes detailed in this proposal follow a few principles specific to arithmetic representation to help ensure a consistent and discoverable user experience for developers working with the Quantum Development Kit.} (QuantumLibraries issue \#337)                     
                                                                                                                 &  3                                    & 0.71\%             \\   
\bottomrule 
\end{tabular}%
\end{table}}

\textbf{(1) Maintainability}: \textit{Maintainability} refers to the ease with which quantum software systems can be modified to correct faults, improve performance, or adapt to evolving requirements. This quality attribute is crucial given the rapid evolution of quantum technologies, where software must be structured in a way that accommodates frequent updates and enhancements. Achieving high maintainability requires clear, modular architecture and comprehensive documentation to facilitate future updates, testing, and debugging in the inherently complex quantum computing landscape. In our dataset, developers mentioned \textit{Maintainability} quality attribute in about 18.53\% of the quality attributes we collected.

\textbf{(2) Performance}: \textit{Performance} plays a pivotal role as it directly influences the computational power and efficiency of quantum algorithms. Quantum computers are designed to excel in specific computational tasks, and optimizing performance ensures that these systems can execute complex quantum algorithms with minimal execution time, making them viable for practical applications such as cryptography, optimization, and material science simulations. In our dataset, developers mentioned \textit{Performance} quality attribute in about 15.68\% of the quality attributes we collected.

\textbf{(3) Compatibility}: \textit{Compatibility} ensures that quantum software systems can interact with classical systems, enabling hybrid quantum-classical computing approaches. This quality attribute is crucial for practical quantum adoption, as quantum computers are often used in conjunction with classical systems for solving complex problems. In our dataset, developers mentioned \textit{Compatibility} quality attribute in about 14.25\% of the quality attributes we collected.

\textbf{(4) Consistency}: \textit{Consistency} plays an important role in quantum software systems, ensuring that quantum computations produce reliable and reproducible results. Achieving consistent outcomes is essential in quantum computing, as it guarantees the correctness of quantum algorithms and maintains data integrity throughout complex quantum operations. Quantum systems must exhibit a high degree of internal and external consistency to meet the rigorous demands of quantum computing tasks. In our dataset, developers mentioned \textit{Consistency} quality attribute in about 8.08\% of the quality attributes we collected.

\textbf{(5) Complexity}: \textit{Complexity} in quantum software systems refers to the intricacy and difficulty in ensuring scalability, efficiency, and robust quantum software design. Balancing these factors demands a profound understanding of both quantum mechanics and software structures, emphasizing the need for advanced tools and methodologies to navigate the technology landscape. In our dataset, developers mentioned \textit{Complexity} quality attribute in about 7.84\% of the quality attributes we collected.

\textbf{(6) Flexibility}: \textit{Flexibility} in quantum software systems ensures that the systems can adapt to varied computational tasks, accommodating different quantum algorithms and protocols. It is vital for catering to a broad spectrum of quantum applications. In our dataset, developers mentioned \textit{Flexibility} quality attribute in about 5.94\% of the quality attributes we collected.

\textbf{(7) Usability}: \textit{Usability} is a key consideration in quantum software systems to facilitate the effective utilization of quantum resources by developers and end-users. User-friendly interfaces, tools, and documentation are essential for enabling practitioners to harness the power of quantum computing without requiring an in-depth understanding of quantum mechanics. In our dataset, developers mentioned \textit{Usability} quality attribute in about 4.99\% of the quality attributes we collected.

\begin{tcolorbox}[colback=gray!5!white,colframe=gray!75!black,title=Key Findings of RQ4]
\textbf{Finding 4}: 

We identified 19 categories of quality attributes that are considered by developers when making architectural decisions in quantum software systems, with \textit{Maintainability} (18.53\%), \textit{Performance} (15.68\%), and \textit{Compatibility} (14.25\%) being the most considered attributes by developers (see more details with examples
in Table \ref{qualityAttribute}).
\end{tcolorbox}

\subsection{RQ5. What are the limitations and challenges of making architecture decisions in quantum software development?} 
\label{resultsofRQ5}

To answer RQ5, we collected and analyzed a total of 343 instances of limitations and challenges (320 from GitHub issues and 41 from Stack Exchange posts). During this process, we found that one GitHub issue and Stack Exchange post may discuss multiple limitations and challenges related to architecture decision-making in quantum software development. If we found the same limitations and challenges within the same post or issue, we counted them only once. We finally identified 16 types of limitations and challenges that developers encounter when making architectural decisions in quantum software development, as presented in Fig. \ref{ResultFigure}e and detailed in Table \ref{limitationAndChallenges}. 

We identified two main types of limitations and challenges: \textit{Design Issues} and \textit{High Error Rates} which practitioners encountered in 25.07\% and 11.37\% of the limitations and challenges we collected. Other types were mentioned in less than 10\% of the limitations and challenges we collected. A detailed description of the top 10 limitations and challenges is provided below.

{\renewcommand{\arraystretch}{1}
\begin{table} [!]
\small
\centering  
\caption{Limitations and challenges of making architecture decisions in quantum software development, their counts \& percentages}
\label{limitationAndChallenges}
\begin{tabular}{p{3cm}p{8.2cm}p{1cm}p{1cm}}
\toprule
\textbf{Limitations \& Challenges}            &  \textbf{Example}           & \textbf{Count}           & \textbf{\%} \\ \hline                
Design Issue                    & \textit{A CurveManager SoftwareModule replacing pyinstruments would definitely fit very nicely in this modular architecture. In this case, there are a lot of design mistakes that we could avoid with our experience of pyinstruments.} (PyRPL issue \#84)                                                     
                                                                                & 86                            & 25.07\%       \\ \hline
High Error Rates                &  \textit{The build systems has also only seen compilation on half a dozen architectures and setups, you may encounter errors in the build process.}   (psi4 \#468)                                                       
                                                                                & 39                           & 11.37\%             \\ \hline
Performance Issue               & \textit{The architecture implemented at the moment is showing some issues that are non trivial to fix: pickling, monitoring, performance, in general seems to hard to grasp for the users intermittent test failures due to syncing issues.} (QCoDeS issue \#272)                                                
                                                                                & 32                            & 9.33\%      \\ \hline
                                                                                
Implementation Issue            & \textit{Currently the API for KEX makes it a bit difficult to implement a KEM transform like Fujisaki-Okamoto generically. One solution (to do it generically) that would involve a somewhat major architectural change is to implement a generic interface.} (liboqs issue \#50)                                                                        
                                                                                & 25                            & 7.29\%         \\ \hline 
Complexity Issue                & \textit{While this architecture is fine for the first ARTIQ systems that were rather small and simple and even for single-crate Metlino/Sayma systems, it shows limitations on more complex ones.} (ARTIQ issue \#778)                  
                                                                                & 24                            & 7.00\%       \\ \hline 
Code Issue                      & \textit{Just reproducing the first issue with code that is fully executable: (..). I think the issue here is purely to do with labelling. When specifying subsystem\_list=[1], DynamicsBackend is only aware that it has a system labelled by index 1 (and it will internally automatically build some measurement definitions based on this labelling).} (qiskit-dynamics issue \#235)               
                                                                                & 22                            & 6.41\%      \\ \hline 
Bugs                            & \textit{Bugs: there are a variety of bugs in the current compiler which have been known about for a long time.} (ARTIQ issue \#1542)                                                                                                
                                                                                & 20                            & 5.83\%          \\ \hline
Compatibility Issue             & \textit{At the same time, we need to plan a transition that doesn't entirely break backwards compatibility.} (deepchem issue \#1039)                                                                                                  
                                                                                & 16                            & 4.66\%        \\ \hline   
Data Issue                      & \textit{Problem is my data consists of 2 arrays consisting of Double numbers representing the real and imaginary parts of complex numbers.} (SO post \#55763054)                                                                     
                                                                                & 16                            & 4.66\%          \\ \hline
Standardization Issue       & \textit{But until/unless NIST gives a standardized version of that, we plan to stick with just the KEM formulation.} (qmcpack issue \#649)                                                                                          
                                                                                & 16                            & 4.66\%          \\ \hline
Maintenance Issue               & \textit{Supporting it would mean a more confused internal architecture, which would hurt us in maintainability in the long term;} (Qiskit issue \#1852)                                                                            
                                                                                & 13                            & 3.79\%         \\ \hline
Usability Issue                 & \textit{To be honest the BackendConfiguration class is not the most ergonomic to use.} (QC post \#28416)               
                                                                                & 11                            & 3.21\%          \\ \hline  
Integration Issue               & \textit{Integration with Keras: Although they're very similar, TensorGraph layers have certain features that it isn't obvious to me how to reproduce with Keras layers.} (deepchem issue \#1038)                              
                                                                                & 8                             & 2.33\%            \\ \hline 
Security Issue                  & \textit{The pseudorandom number generator used by random.random(), Mersenne Twister, is not suitable for cryptography or information security.} (SO post \#63609099)
                                                                                & 7                             & 2.04\%          \\ \hline

Lack of Effort                  &  \textit{A new interface would provide the most benefit here, but it would also require the most effort, both in development and long-term maintenance.} (qsim issue \#293)                                                        
                                                                                & 5                             & 1.46\%             \\ \hline 
Noise Issue                  &  \textit{This feature will enhance the current PyQuil simulator's capabilities (noise.py module) by providing a more accurate representation of noise.}(pyquil issue \#1575)                                                        
                                                                                & 3                             & 0.87\%            \\ 

\bottomrule 
\end{tabular}%
\end{table}}

\textbf{(1) Design Issues}: \textit{Design Issues} are key challenges that developers encounter most when making architecture decisions in quantum software development. These challenges arise when developing quantum algorithms, optimizing quantum circuits, and seamlessly integrating them with components. Decision-making on design issues is particularly challenging due to the evolving standards and tools in quantum computing, which require learning and adaptation. To address \textit{Design Issues}, it is important for developers to first learn about quantum principles and use quantum algorithms. Developers also need to consider the specific hardware and software stack being used. In our dataset, developers discussed \textit{Design Issues} in about 25.07\% of the limitations and challenges we collected.

\textbf{(2) High Error Rates}: \textit{High Error Rates} in quantum software architecture refer to the frequent occurrence of errors during the execution of quantum algorithms, compilation, or deployment of quantum software systems.  High error rates in quantum software architecture are a significant challenge arising from various sources, including asynchronous operations, build and compilation issues, inadequate error mitigation techniques, and incorrect configurations. Addressing these errors involves improving synchronization, enhancing error mitigation techniques, ensuring cross-platform compatibility, correctly implementing quantum algorithms, managing configurations properly, and conducting thorough testing. Developers can reduce error rates and build more robust quantum software systems by focusing on these areas. In our dataset, developers discussed \textit{High Error Rates} in about 11.37\% of the limitations and challenges we collected.

\textbf{(3) Performance Issues}: \textit{Performance Issues} in quantum software arise when applications do not fully leverage quantum speedups, often due to inefficient algorithms or sub-optimal resource utilization. Developers face these challenges when making architectural decisions because identifying and using quantum advantages requires a deep understanding of both quantum mechanics and the problem domain. To prevent \textit{performance Issues}, developers need to focus on algorithm optimization specific to quantum processing, ensuring that quantum resources are effectively utilized to offer a true computational advantage. In our dataset, developers discussed \textit{Performance Issues} in about 9.33\% of the limitations and challenges we collected.

\textbf{(4) Implementation Issues}: \textit{Implementation Issues} in quantum software development primarily stem from the intricate nature of quantum algorithms and the complexities inherent to nascent technologies, as illustrated by challenges in integrating quantum algorithms with existing classical computing systems. Specifically, developers often encounter difficulties when trying to generalize functions or leverage advanced features due to hardware-software co-optimization, and practical and robust frameworks that can seamlessly integrate with various quantum algorithms and protocols. These challenges are magnified by the high complexity of quantum computations, nuances of post-quantum cryptography, and the infancy of quantum programming languages. To prevent these issues, developers should focus on fostering adaptive software designs, ensuring rigorous testing and validation of newer features, deepening collaboration between quantum software communities, and enhancing the flexibility and documentation of quantum development tools and APIs. In our dataset, developers discussed \textit{Implementation Issues} in about 7.29\% of the limitations and challenges we collected.

\textbf{(5) Complexity Issues}: Quantum software inherently presents \textit{Complexity Issues} due to the multi-faceted nature of quantum mechanics. Developers wrestle with this type of issue when making architecture decisions, as understanding and managing the intricacies of superposition and entanglement can be daunting. This complexity makes designing efficient quantum algorithms challenging. To alleviate \textit{Complexity Issues}, a deep dive into quantum principles, alongside leveraging abstraction layers and tools that simplify quantum processes, is indispensable. In our dataset, developers discussed \textit{Complexity Issues} in about 7.00\% of the limitations and challenges we collected.

\textbf{(6) Code Issues}: \textit{Code Issues} in quantum software involve problems like unreadable code, lack of proper documentation, and non-reusability, impeding the functionality or efficiency of applications. This type of issue is prevalent due to the developing stage of quantum programming languages and the complex nature of quantum computations. To mitigate code issues, developers should adhere to best coding practices, including thorough documentation, code reviews, and leveraging classical coding standards where applicable. In our dataset, developers discussed \textit{Code Issues} in about 6.41\% of the limitations and challenges we collected.

\textbf{(7) Bugs}: In quantum software systems, \textit{Bugs} refer to inconsistencies, or unintended behaviors within the software that hinder its correct or efficient operation. \textit{Bugs} can arise from a multitude of sources, including long-standing compiler issues, incorrect dataset creation, and improper support for certain measurements due to runtime updates. Developers can address these issues by conducting thorough testing, profiling for resource management, and ensuring backward compatibility with runtime updates. Addressing \textit{Bugs} also involves refining unit tests to reflect actual usage patterns, consolidating coding styles, and improving error messages to reduce user confusion. Regular profiling and debugging, especially for resource-heavy operations, are crucial in identifying and mitigating these bugs, ultimately enhancing the reliability and user experience of quantum software systems. \textit{Bugs} accounts for 5.83 \% in our analysis of quantum software systems.

\textbf{(8) Compatibility Issues}: \textit{Compatibility Issues} in quantum software development are primarily related to the challenges of creating compatible interfaces between quantum and classical systems. These difficulties are compounded by the absence of universally accepted quantum programming frameworks and the significant obstacles in integrating quantum technology with existing IT infrastructures. Together, these factors pose substantial challenges to the advancement and application of quantum computing in current technological environments. This incompatibility extends to difficulties in data formats, communication protocols, and system interoperations, which underscore an essential need for enhanced functionalities across different layers of the software stack as well as improved interfaces between quantum and classical computing environments. In our dataset, developers discussed \textit{Compatibility Issues} in about 4.66\% of the limitations and challenges we collected.

\textbf{(9) Data Issues}: \textit{Data Issues} refer to challenges in data encoding, storage, and retrieval using quantum bits in quantum computing. Developers encounter this type of issue when making architecture decisions because the intricate process of encoding classical information into qubits is influenced by the nature and encoding type of data, which can significantly impact the performance of algorithms. To counteract \textit{Data Issues}, developers should choose the appropriate qubit data encoding patterns, understand quantum data operations, and utilize appropriate quantum data structures. In our dataset, developers discussed \textit{Data Issues} in about 4.66\% of the limitations and challenges we collected.

\textbf{(10) Standardization Issues}: \textit{Standardization Issues} affect quantum software development, as the field suffers from a lack of standardized tools and frameworks. Encouraging collaborative efforts towards standardizing quantum computing protocols and interfaces, and contributing to community-driven quantum software libraries can help establish much-needed uniformity. In our dataset, developers discussed \textit{Standardization Issues} in about 4.66\% of the limitations and challenges we collected.

\begin{tcolorbox}[colback=gray!5!white,colframe=gray!75!black,title=Key Findings of RQ5]
\textbf{Finding 5}:

\textit{Design Issues} (25.07\%) and \textit{High Error Rates} (11.37\%) are the most common challenges in architecture decision-making in the development of quantum software systems. Other notable issues include \textit{Performance Issues} (9.33\%), \textit{Implementation Issues} (7.29\%), \textit{Complexity Issues} (7.00\%) and \textit{Code Issues} (6.12\%) (see the examples in Table \ref{limitationAndChallenges}). 

\end{tcolorbox}

\section{Discussion}\label{discussionAndImplications}
In this section, we revisit the findings of this study by interpreting the results in Section~\ref{analysisoftheresults} and discussing their implications for researchers
(indicated with the {\faLeanpub \hspace{0.5mm}} icon) and/or practitioners (the \faMale \hspace{0.5mm} icon) in Section~\ref{implications}. 

\subsection{Interpretation of results}\label{analysisoftheresults}
\subsubsection{RQ1: How do developers express architecture decisions made in quantum software development?}
Table \ref{linguisticPatterns} presents the linguistic patterns identified in the architecture decisions made during the development of quantum software systems. These patterns serve as a means to understand the diverse intentions behind decision expressions, shedding light on the reasoning process that drives the architects and developers of quantum software systems. Among these linguistic patterns, two arise as particularly dominant: \textit{Solution Proposal} and \textit{Information Giving}. \textit{Solution Proposal} is crucial for suggesting multiple solutions to architectural challenges in quantum technologies. Its frequent use underscores its importance in enriching decision-making processes by exploring alternative pathways. \textit{Information Giving} plays a critical role in quantum software development by conveying essential context and insights necessary for informed decision-making. It facilitates effective communication among architects and developers, ensuring a comprehensive understanding of technical details. While Solution Proposal and Information Giving are prominent,\textit{Feature Request}, \textit{Problem Discovery}, and \textit{Opinion Asking} are less frequently utilized in expressing architecture decisions in quantum software development. These patterns offer potential avenues for enhancing collaborative decision-making but require further exploration within this context.

%\textit{Solution Proposal}, as the name suggests, plays a pivotal role in suggesting potential solutions as architecture decisions. Given the multifaceted nature of quantum technologies, any given architectural challenge may have a multitude of viable solutions. In such cases, practitioners lean on the \textit{Solution Proposal} pattern to put forth various potential solutions, thereby enriching the decision-making process. \textit{Information Giving}, on the other hand, plays a critical role in the landscape of quantum software development. When architects and developers grapple with complex quantum software development challenges, they often need to convey pertinent information to elucidate the intricacies of their decision-making. This is where \textit{Information Giving} comes into play, enabling professionals to share the necessary context and insights required for effective architecture decision-making in the quantum computing domain. In essence, the quantum technology landscape is marked by its dynamism and diversity, which inherently gives rise to a multitude of potential solutions for architectural problems. Hence, \textit{Solution Proposal} acts as a common linguistic pattern for quantum practitioners to explore and propose alternative pathways, and \textit{Information Giving} provides an essential bridge for communication and understanding of technical details, facilitating the exchange of vital information necessary for sound architecture decision-making within the intricate field of quantum software development.

\subsubsection{RQ2. What types of architecture decisions are made in quantum software development?}
Our investigation indicates the presence of 15 types of architecture decisions in the context of quantum software development, as shown in Table \ref{typeOfArchitectureDecision}. The prominent  \textit{Implementation Decision} underscores the necessity for detailed and precise implementation guidelines in quantum systems. This emphasizes the importance of thoroughly understanding how to integrate and utilize quantum components, given their complexity and the current nascent stage of quantum technology. Similarly, \textit{Technology Decision} is critical due to the rapid evolution of quantum technologies. Practitioners must navigate a rapidly changing landscape to select appropriate technologies that align with their specific system requirements. For instance, selecting a quantum programming framework or backend configuration can significantly impact system performance and maintainability, as illustrated by the ongoing developments in tools like Qiskit and OpenQL. \textit{Architecture Pattern Decision} guides the selection of suitable architectural patterns to meet specific quality attributes like maintainability and performance. This decision is crucial for ensuring that quantum systems can adapt and evolve over time, accommodating new features and optimizations without significant rework. \textit{Component Decision}, which emphasizes modular and layered architectures, facilitates maintainability and scalability, enabling developers to manage complex systems more effectively. \textit{Tool Decision} and \textit{Data Decision} highlight the importance of selecting optimal tools and managing data efficiently. Given the significant and dynamic nature of data generated during quantum computing tasks, these decisions are pivotal for ensuring robust and efficient data handling and processing. Other architecture decisions within quantum systems encompass \textit{Decomposition Decision}.\textit{Integration Decision} focuses on melding quantum components with traditional systems for smooth data and process interactions. \textit{Behavior Decision}, \textit{Design Pattern Decision}, \textit{Service Decision}, \textit{Testing Decision}, \textit{Deployment Decision}, \textit{Containerization Decision} and \textit{Concurrency Decision}. Collectively, these decisions underscore the multifaceted considerations that architects and developers need to address while designing quantum software systems.

\subsubsection{RQ3. What are the application domains of quantum software in which architecture decisions are made?} 
\label{interpretationOfRQ3Results}

As illustrated in Table \ref{domainArchitectureDecisions}, architecture decisions in quantum software development are made in a total of 20 application domains, in which \textit{Software Development Tools} are dominant. The prominence of \textit{Software Development Tools} is attributed to their critical role in accessing and managing quantum-specific modules and functionalities required for quantum software systems. These tools facilitate the development, integration, and optimization of quantum algorithms and applications, leveraging frameworks that bridge classical and quantum computing paradigms. Quantum-specific software development environments, such as those provided by Qiskit, Cirq, and PennyLane, enable developers to implement quantum algorithms, manage quantum states, and interface with quantum hardware effectively. Beyond \textit{Software Development Tools}, \textit{Quantum Programming Framework} plays a pivotal role in quantum software architecture decisions, particularly in defining the foundational structures and methodologies for quantum algorithm implementation. These frameworks encompass libraries, compilers, and runtime environments tailored to quantum computing requirements, influencing architecture decisions around algorithm design, optimization techniques, and hardware compatibility.  \textit{Scientific Software} harnessing the power of quantum mechanics, and \textit{Optimization and Simulation} solutions that leverage principles of quantum mechanics. Quantum phenomena also play a pivotal role in domains such as \textit{Quantum Cryptography}, offering unmatched security features. Likewise, the synergy between quantum mechanics and artificial intelligence is evident in the \textit{Deep Learning Framework} domain, where quantum-enhanced neural networks are under exploration. Quantum technology is causing significant changes (e.g., optimized quantum algorithms, enhanced computational power, and increased data processing capabilities) in domains like \textit{Quantum Compiler}, \textit{Real-Time System}, \textit{Distributed System}, and \textit{Embedded System}. Moreover, \textit{Blockchain Technology} is witnessing innovative approaches influenced by quantum mechanics, ensuring robust and tamper-proof systems. Traditional \textit{Desktop Applications}, \textit{Web Applications}, and \textit{Mobile Applications} are also experiencing a transformative phase, with quantum algorithms enhancing their efficiency and capabilities. Finally, \textit{Games} in quantum software systems are still in their nascent stages, but hold the promise of offering incomparable gaming experiences, tapping into the probabilistic nature of quantum mechanics.

\subsubsection{RQ4. What quality attributes are considered when developers make architecture decisions in quantum software development?} 
As shown in Table \ref{qualityAttribute}, we identified 19 types of quality attributes considered by architects when they make architecture decisions in quantum software development. In our study, two prominent quality attributes, \textit{Maintainability} and \textit{Performance} were mostly discussed in quantum software development. \textit{Maintainability} was mentioned 78 times, constituting over 18\% of the identified quality attributes, while \textit{Performance} was found 66 times, accounting for approximately 15\% of the total quality attributes. \textit{Maintainability} involves ensuring that the system remains easy to manage and evolves over time. \textit{Maintainability} plays an essential role in quantum software systems since it ensures that quantum software systems can evolve as technology advances or requirements change. As one developer proposed ``\textit{I think the node should be kept minimal and specific for a lot of good architectural reasons. (...) More non-core features, more on/off settings => more complexity, less maintainability, higher risk}'' (QRL issue \#1545).  \textit{Performance} is another vital quality attribute for quantum software systems, since \textit{Performance} is crucial for long-term viability and adaptability. \textit{Performance} is primarily associated with time behavior, resource utilization, and system capacity. Practitioners in quantum software development are particularly concerned about factors such as training and computing time, as well as the speed of data processing. These aspects are crucial for optimizing the performance of quantum software systems, and suggestions include the adoption of better-distributed architectures to enhance these performance aspects \cite{koziolek2014automated}, as mentioned in a GitHub issue ``\textit{First the current design is intrinsically bad performance for hot loops because we have to look up the state (even if it doesn't change)}'' (DFTK.jl issue \#313).  

\subsubsection{RQ5. What are the limitations and challenges of making architecture decisions in quantum software development?} 
As shown in Table \ref{limitationAndChallenges}, developers encounter 16 types of limitations and challenges during architecture decision-making of quantum software development, in which \textit{High Error Rates} and \textit{Design Issues} are the main types. \textit{Design Issues} frequently arise in quantum software development due to e.g., complex quantum technology, quantum software development tools, and quantum programming frameworks. Quantum software not only requires a redesign of applications, but also mandates a reconceptualization of the tools and languages used for quantum software development. Furthermore, from a coding perspective, there are hurdles to overcome to maintain clarity and scalability in the quantum context, as highlighted in an issue that states, ``\textit{This issue is meant to highlight some design issues from the coding point of view that will make life hard for current and future coders of RMG-Py}'' (RMG-Py issue \#521). \textit{High Error Rates}, on the other hand, are tightly intertwined with the intrinsic characteristics of quantum systems. Given the delicate nature of quantum states and the probabilistic outcomes they produce, ensuring consistent and efficient performance is a challenge. For instance, a developer noted: ``\textit{The current design is an intrinsically bad performance for hot loops because we have to look up the state (even if it doesn't change)}'' (DFTK.jl issue \#313). This example touches upon the nuances of state management in quantum software systems, where even seemingly minor operations can entail significant overheads. In another context, as quantum algorithms are integrated into larger systems, unpredictable performance hiccups can occur, as evidenced by the feedback from one developer that ``\textit{But the performance got slightly worse (higher variance) or the model crashed due to NaNs}'' (deepqmc issue \#25).

\subsection{Implications} \label{implications}
\subsubsection{Linguistic patterns}
\textbf{ Identifying linguistic patterns helps researchers understand architecture decisions decision description, while practitioners can use these insights to improve architecture communication in quantum software projects:} The identification of linguistic patterns (e.g., \textit{Solution Proposal}, \textit{Information Giving}, \textit{Information Seeking}, and less commonly used \textit{Feature Request}, \textit{Problem Discovery}, and \textit{Opinion Asking}) provides a structured framework for understanding decision-making processes in quantum software architecture. \faLeanpub \hspace{0.5mm} For researchers, linguistic patterns enhance the study of decision-making dynamics, offering insights that can inform the development of improved decision-support tools and methodologies tailored to quantum software architecture. By analyzing these linguistic patterns, researchers gain deeper insights into how architectural decisions are formulated, communicated, and evolved within quantum computing environments. This foundational knowledge can lead to theoretical advancements in decision-making models, contributing significantly to the broader field of software engineering and quantum computing research. \faMale \hspace{0.5mm} For industry practitioners, understanding and leveraging these patterns fosters better collaboration and communication on architecture among teams in quantum software development projects. The dominance of \textit{Solution Proposal} and \textit{Information Giving} underscores the importance of fostering flexible and communicative decision-making frameworks within industry settings. \faMale \hspace{0.5mm} Practitioners can utilize these insights to streamline decision processes, improve the quality of architectural solutions, and effectively manage project complexities. \textit{Information Seeking} though less frequent, plays a critical role in seeking clarifications and resolving technical uncertainties, which is essential for optimizing decision-making efficiency. \faMale \hspace{0.5mm} Practically, integrating features into architecture tools and platforms that support \textit{Solution Proposal}, \textit{Information Giving}, and \textit{Information Seeking} can lead to the development of advanced Integrated Development Environments (IDEs) and collaborative platforms tailored to the needs of quantum software practitioners. %The insights derived from this study are \faLeanpub \hspace{0.5mm} instrumental in designing educational frameworks aimed at \faMale \hspace{0.5mm} training future quantum software practitioners. Emphasizing these linguistic patterns in educational programs equips graduates with essential skills such as proposing solutions, sharing information comprehensively, and collaborating effectively within interdisciplinary teams.

\textbf{\textit{Feature Request}, \textit{Problem Discovery}, and \textit{Opinion Asking} can be considered by architects and developers in describing their architecture decisions in quantum software development:} To advance quantum software architecture effectively, \textit{Feature Request}, \textit{Problem Discovery}, and \textit{Opinion Asking} were found to be less frequently used for expressing architecture decisions during the development of quantum software. \textit{Feature Request} typically outlines a new functionality or enhancement being added to the system. The given example, ``\textit{Pull requests implementing new features: Data\-layer type creation, conversion and dispatch \#1338 implementing the data-layer creation, conversion and dispatcher routines}'' (QUTIP issues \#1278), specifies a technical enhancement, indicating that pull requests have been made for new data layer functionalities, which are crucial for the data handling capabilities of software systems. Furthermore, \textit{Problem Discovery} is about identifying and describing issues or limitations within the current architecture. The example ``\textit{SIKE not properly disabled on unsupported architectures}'' (liboqs issue \#1128) highlights a specific issue where a cryptography algorithm SIKE is not being correctly disabled in environments where SIKE is not supported, potentially leading to errors or security vulnerabilities. Lastly, \textit{Opinion Asking} is a linguistic pattern where developers seek feedback or alternative perspectives on an architecture decision, promoting collaborative decision-making. The example ``\textit{Please feel free to suggest changes or alternate designs}'' (deepchem issue \#2059) opens the floor for community input on the project, which can lead to innovative solutions and shared ownership of the project's evolution. {\faLeanpub \hspace{0.5mm}} Researchers and  \faMale \hspace{0.5mm} practitioners are encouraged to integrate these patterns into their decision-making processes to enhance the agility, robustness, and innovation of quantum software architectures.

\subsubsection{Architecture decisions}
\textbf{\textit{Implementation Decisions} and \textit{Technology Decisions} are critical architecture decisions, requiring deep understanding and strategic planning in quantum software development:} The identified decisions directly influence the structure and functionality of quantum software architectures. Prominent among these decisions are \textit{Implementation Decision} and \textit{Technology Decision}, the complexity of which is underlined by the need to manage qubit. The intricacies of these decisions require a deep understanding of quantum components and their integration, contributing to more mature and robust architecture designs. These decisions do not merely guide the development process; they also ensure that the architecture can handle the unique requirements of quantum computing. For instance, the architecture decision-making process becomes intricately complex. {\faLeanpub \hspace{0.5mm}} Practitioners often find themselves at a crossroad, deliberating on the most suitable means to incorporate quantum-related components. Such as, determining the feasibility of introducing a feature, like caching, within a given architecture poses questions related to effort, compatibility, and user comprehension, as evidenced in this example ``\textit{The C++ API of LLVM is complicated and has poor forward compatibility (...). I think we all agree that caching is a good idea; the question is just how easy it is to implement in the current architecture (in terms of implementation effort, as well as ease of understanding for users)}'' (ARTIQ issue \#1542). On the other hand, \textit{Technology Decision} delves into the selection and integration of appropriate technologies that cater to the needs of quantum software systems. {\faLeanpub \hspace{0.5mm}} The rapid evolution of quantum technologies and tools bewilders many practitioners. A case in point is the quandary expressed by a developer, ``\textit{What's the latest in Python LLVM bindings anyway \- is there one that supports the (C++) IRBuilder interface?}'' (ARTIQ issue \#1370), where the binding specifics for a language (Python) with a compiler (LLVM) are in question.

\textbf{Researchers should innovate methodologies to address challenges, while practitioners need modular, layered approaches to handle evolving quantum technology:} For the research community, these findings highlight key areas that require further exploration and innovation. {\faLeanpub \hspace{0.5mm}} Researchers can focus on developing new methodologies and frameworks to address the identified challenges, such as creating more effective architecture patterns and optimization techniques for quantum software systems. This can lead to significant advancements in the field of quantum software engineering. Therefore, we echo the need for {\faLeanpub \hspace{0.5mm}} more empirical studies to fully understand and quantify the benefits of these insights for quantum software architecture.  \faMale \hspace{0.5mm} For industrial practitioners, \textit{Implementation Decisions} and \textit{Technology Decisions} underscore the importance of strategic planning and continuous learning. The findings also impact design practices, encouraging practitioners to adopt a modular and layered approach to architecture design. This approach enhances the flexibility and adaptability of quantum software systems, allowing for easier updates and integration of new technologies. \faMale \hspace{0.5mm} Practitioners were also faced with significant challenges when making \textit{Implementation Decisions} and \textit{Technology Decisions} during the development of quantum software, particularly due to the integration of quantum-related components and the intricate nature of quantum technology, which often leads to critical issues. Consequently, we suggest that practitioners should carefully consider \textit{Implementation Decisions} and \textit{Technology Decisions} in quantum software development.

\subsubsection {Application domains}

\textbf{ Researchers should focus on refining integration, developing modular designs, and implementing advanced error mitigation for reliable quantum software:} Our findings have several key implications for both the academic community and industry professionals. {\faLeanpub \hspace{0.5mm}} Researchers should focus on refining the integration procedures between various quantum programming frameworks and tools. Ensuring seamless compatibility between different modules and APIs is critical. For example, ``\textit{Align current BaseJob API with the specification}'' (Qiskit issue \#933) and ``\textit{The SDK also provides the ability to request specific result types on a subset of qubits}'' (amazon-braket-sdk-python issue \#158) illustrate the need for continuous improvement in integration to support diverse quantum computing tasks. Emphasis should be placed on the design and evolution of modular quantum software architectures. The decomposition and refactoring of components, such as ``\textit{After we have finished the decomposition and local Covalent has a microservice architecture}'' (covalent issue \#186), highlight the importance of maintaining flexibility and scalability in quantum software systems. {\faLeanpub \hspace{0.5mm}} Researchers should explore innovative modular designs to accommodate the evolving requirements of quantum computing. The development of robust error mitigation techniques, as indicated by ``\textit{Mitiq needs more error mitigation techniques}'' (mitiq issue \#821), is crucial for improving the reliability and accuracy of quantum computations. {\faLeanpub \hspace{0.5mm}} Researchers should investigate and implement advanced error correction methods to enhance the performance of quantum algorithms. The rapid evolution of quantum computing paradigms necessitates adaptive and forward-thinking research approaches. Proposals like ``\textit{This issue is motivated by problems emerged in \#471 (Bayesian ZNE), \#477 (Clifford-data regression)}'' (mitiq issue \#517) underscores the need for continuous adaptation to emerging quantum computing challenges and opportunities.

\textbf{Practitioners must ensure tool compatibility, create user-friendly interfaces, and adopt rigorous testing and modern development practices for enhanced quantum software:} \faMale \hspace{0.5mm} Practitioners must ensure the compatibility and coherence of tools and interfaces used in quantum software development. Issues like ``\textit{The sync method in covalent/\_results\_manager/results\_manager.py needs to be moved and updated to be compatible with the new covalent services architecture}'' (covalent issue \#370) highlight the necessity for ongoing maintenance and updates to ensure smooth operations across different components. Developing user-friendly and flexible tools is essential for the practical implementation of quantum software. Statements such as ``\textit{We are also interested in using the Python SDK}'' (Qiskit issue \#487) and ''\textit{While the actual operation requested is technically possible without going to a DAG, it doesn't fit into the compilation tooling architecture of Terra}'' (Qiskit issue \#1852) emphasize the need for designing tools that are both technically robust and accessible to users. \faMale \hspace{0.5mm} Practitioners should adopt rigorous testing and continuous improvement practices. Examples like ``\textit{Fix unit tests (locally there is no problem, but travis keeps on failing due to connection issues) modify docs}'' (pyrpl issue \#350) and ``\textit{Adapt coverage testing to require 100\% of all new code}'' (circuit-knitting-toolbox issue \#107) illustrate the importance of thorough testing and validation to ensure the reliability of quantum software systems. The adoption of modern development practices, such as microservices~\cite{ahmad2023engineering} and agile methodologies~\cite{yang2016systematic}, can significantly enhance the development process. The shift towards architectures like ``\textit{local Covalent has a microservice architecture}'' (covalent issue \#186) demonstrates the benefits of modular, scalable, and maintainable software architecture designs in quantum system development.

\subsubsection{Quality attributes}
\textbf{The quality attributes considered when making architecture decisions in quantum software development offer valuable insights for both researchers and practitioners:} The quality attributes delve into the crucial aspects of software architecture in quantum software development. {\faLeanpub \hspace{0.5mm}} Understanding these quality attributes is essential for developing efficient and maintainable quantum software systems, guiding both the research agenda and practical approaches in quantum software development. Therefore, there is a need to provide guidelines for both {\faLeanpub \hspace{0.5mm}} researcher and \faMale \hspace{0.5mm} practitioners. {\faLeanpub \hspace{0.5mm}} Researchers can explore the ways in which quality attributes are considered when formulating architectural decisions in quantum software development. {\faLeanpub \hspace{0.5mm}} Researchers can also use this knowledge to focus on the identified quality attributes (e.g., \textit{Maintainability}, \textit{Performance}, and \textit{Compatibility}), tailoring their theoretical frameworks and empirical studies to address the specific challenges and needs in quantum software architecture. Recognition of the importance of quality attributes in quantum software development can be directed towards constructing methodologies and tools that enhance these quality attributes in quantum software systems. This focus is crucial, given the inherent complexities of quantum software and the need for integrating quality attributes in quantum software with classical software paradigms.
\faMale \hspace{0.5mm} Practitioners can utilize our findings of quality attributes to make well-informed architecture decisions during the development of quantum software. The emphasis on \textit{Maintainability}, \textit{Performance}, and \textit{Compatibility} as key quality attributes indicates the areas where \faMale \hspace{0.5mm} practitioners should focus their efforts. This knowledge assists in creating more robust and efficient quantum software systems, directly impacting the practical application of quantum technologies. Furthermore, understanding these quality attributes can aid in making valid architecture decisions, ensuring that quantum software not only meets technological demands but is also maintainable and compatible to future advancements and requirements.

\subsubsection{Limitations and challenges}

\textbf{Practitioners and researchers face significant \textit{Design Issues} in quantum software development, requiring adaptations of traditional practices and exploration of new frameworks like layered software architecture:} \faMale \hspace{0.5mm} Practitioners and {\faLeanpub \hspace{0.5mm}} researchers alike must grapple with critical challenges and limitations when making architecture decisions in quantum software development. These challenges encompass various aspects, prominently including \textit{Design Issues} that arise from the unique demands of quantum computing, as mentioned by a developer, ``\textit{the authors do lay out some of the challenges faced when designing quantum computing design tools and programming languages}'' (CS post \#29032); this indicates recognition of difficulties in adapting traditional software design practices to the quantum realm and marks the exploration of \faMale \hspace{0.5mm} practitioners into the field of quantum software, focusing on the technical and conceptual challenges in designing effective quantum programming languages and design tools. The proposed solution involves adopting a layered software architecture framework, as mentioned by a developer, ``\textit{a Layered Software Architecture for Quantum Computing Design Tools by Svore et al. lays out a framework that I believe cuts across what you're getting it}'' (CS post \#29032). {\faLeanpub \hspace{0.5mm}} For researchers, understanding and addressing these challenges pave the way for advancing quantum software engineering, and the exploration of quantum software design tools and programming languages marks an important frontier in quantum computing research.
%Poor design decisions in quantum software can lead to complex, inefficient, or unreliable outcomes. For instance, ``\textit{VibrationalOp hard-codes a rather odd and confusing case here because of a poor design choice in the VibrationalIntegrals taken here}'' (qiskit-nature \#1032), indicates how poor design decisions in the early stages can significantly impact the clarity and functionality of the resulting code or system. To mitigate these challenges, two potential solutions were discussed, ``\textit{1) storing the label templates for each key in the PolynomialTensor: 2) somehow attach the label template to the stored array}''  (qiskit-nature \#1032), implementing such solutions can improve the adaptability of quantum software architecture, potentially enabling better support for diverse quantum computing applications and operator types.

\textbf{\textit{High Error Rates} demand advanced error management and innovative error mitigation techniques to improve reliability in quantum software development}: Another prominent challenge identified in this study is the management of \textit{High Error Rates} inherent in quantum software systems. A notable example of this challenge is encountered in quantum chemistry simulations where high error rates can lead to inaccuracies in computational results, as mentioned by a developer, ``\textit{The build systems have only seen compilation on half a dozen architectures and setups, so you may encounter errors in the build process}'' (Psi4 issue \#468). To mitigate this challenge, one potential solution was discussed by a developer, ``\textit{In addition, python-based error messages will be more expressive and accurate. For users who enjoy the current Psithon interface, the current Psi4 Psithon parsing equipment will be moved to a script named ‘psi4’ in the binary location. This will allow all current Psi4 input files to be run normally}'' (Psi4 issue \#468). In this example, python-based error messages can be more expressive and accurate, thereby simplifying the debugging process and addressing \textit{High Error Rates}. Addressing \textit{High Error Rates} challenges requires a dedicated focus on developing advanced error management techniques. Therefore, \faMale \hspace{0.5mm} practitioners should implement enhanced error management capabilities within their quantum software systems. {\faLeanpub \hspace{0.5mm}} Researchers should explore innovative methodologies for error mitigation and correction to improve the reliability of quantum software systems.

%\textcolor{blue}{\textbf{Comprehensive documentation are crucial to avoid failure in quantum software:} To address these architecture challenges and limitations in quantum software development, \faMale \hspace{0.5mm} Practitioners should pay more attention to studying design decisions and choosing suitable technologies. For instance, one developer mentioned, ``\textit{We can resolve this issue with the conclusion `By Design'}'' (qsharp-runtime issue \#587), marks the significance of thoughtful design decisions in overcoming challenges during the development of quantum software. \faMale \hspace{0.5mm} On the other hand, if practitioners are unaware of these limitations and challenges or are unable to address them appropriately, the limitations may exacerbate during the development of quantum systems, potentially leading to a failed quantum software system. As one developer mentioned, ``\textit{The absence of a design document enumerating realistic use cases hides the design goals and blurs the definition of done}'' (Qiskit issue \#825), indicates the critical need for comprehensive design documentation to avoid misalignment of implemented quantum software systems. The importance of making informed architecture decisions in quantum software development cannot be ignored: \faMale \hspace{0.5mm} practitioners awareness and proper handling of the inherent limitations and challenges not only made up the way for successful outcomes but also prevents them from the failures of quantum software systems.}

\textbf{Design documentation is crucial to avoid failure in quantum software:} Both {\faLeanpub \hspace{0.5mm}} researchers and \faMale \hspace{0.5mm} practitioners in the field of quantum software development can benefit significantly from recognizing the importance of design documentation. As highlighted by one developer, ``\textit{the absence of a design document enumerating realistic use cases hides the design goals and blurs the definition of done}'' (Qiskit issue \#825), the absence of a design document that includes realistic use cases can obscure the design goals and make it challenging to define project completion. {\faLeanpub \hspace{0.5mm}} Therefore, it is crucial for researchers to create design documentation that clearly outlines the intended objectives, realistic use cases, and specific requirements of quantum software systems. This design documentation serves as a roadmap for development and helps avoid misalignment between the design goals and the actual implementation of quantum software systems, preventing discrepancies and misunderstandings during the development process. \faMale \hspace{0.5mm} Practitioners can rely on design documentation to understand the intended functionality and requirements of the quantum software system. With well-defined documentation that enumerates realistic use cases, {\faLeanpub \hspace{0.5mm}} researchers and \faMale \hspace{0.5mm} practitioners can better understand the design goals and the definition of project completion. This clarity can lead to more successful outcomes and prevent the failures of quantum software systems.

\subsubsection{Relationship between the RQs}

\textbf{Critical application domains:} The most critical application domains in quantum software in which \textit{Implementation Decisions} are made include: \textit{Software Development Tools}, \textit{Quantum Programming Frameworks}, \textit{Optimization and Simulation}, \textit{Data Processing}, and \textit{Scientific Software}. \textit{Software Development Tools} have the highest number of \textit{Implementation Decisions} (27 out of 125), representing 21.6\%. This indicates a strong focus on tools that support the development process. \textit{Quantum Programming Frameworks} with 23 (18.4\%), this domain emphasizes the importance of frameworks that facilitate quantum programming. \textit{Optimization and Simulation} accounts for 17 \textit{Implementation Decisions} (13.6\%), showing the significance of optimizing and simulating quantum systems. \textit{Scientific Software} with 10 (8\%), this domain highlights the role of scientific applications in quantum computing. \textit{Data Processing} holding 8 (6.4\%), this domain underlines the necessity for efficient data handling in quantum software. These findings underscore the critical roles that \textit{Software Development Tools} and \textit{Quantum Programming Frameworks} play in the architecture of quantum software. \faMale \hspace{0.5mm} Practitioners should focus on selecting and utilizing tools that enhance \textit{Maintainability} and \textit{Performance}, given their significant impact on the success of quantum software applications.

\textbf{Key quality attributes:} The analysis of quality attributes reveals the following insights: \textit{Maintainability} has the highest frequency of \textit{Implementation Decisions} (36 out of 228), accounting for 15.8\%. It highlights the importance of maintaining the software over time. \textit{Performance} with 34 \textit{Implementation Decisions} (14.9\%), performance is a key concern in quantum software development. \textit{Compatibility} has 30 \textit{Implementation Decisions} (13.2\%), emphasizing the need for software to function well with other systems. \textit{Consistency} holding 22 \textit{Implementation Decisions} (9.6\%), consistency is crucial for ensuring reliable and predictable software behavior. \textit{Complexity} with 19 \textit{Implementation Decisions} (8.3\%), managing complexity is essential for developing efficient quantum software. {\faLeanpub \hspace{0.5mm}} These quality attributes (e.g., Maintainability, Performance, Compatibility) highlight critical research areas in quantum software where further research is needed. \faMale \hspace{0.5mm} Understanding these quality attributes is crucial, guiding practitioners to make suitable architectural decisions when developing quantum software systems.

\textbf{Limitations and challenges:} \textit{Design Issues} appeared as the most prevalent challenge, accounting for 48 out of 228 \textit{Implementation Decisions} (21.1\%). This high frequency underscores the common occurrence of design-related challenges in quantum software development. \textit{Performance Issues} with 20 \textit{Implementation Decisions} (8.8\%), performance issues are a significant challenge. \textit{Implementation Issues} accounts for 19 \textit{Implementation Decisions} (8.3\%), indicating difficulties in the implementation process. \textit{Complexity Issues} holding 11 \textit{Implementation Decisions} (4.8\%), complexity remains a notable challenge in quantum software development. \textit{High Error Rate} challenge has 8 \textit{Implementation Decisions} (3.5\%), highlighting the issue of errors in quantum software. Addressing these challenges is crucial for advancing the field and developing robust quantum software systems. The challenges identified, such as \textit{Design Issues} (21.1\%) and \textit{Performance Issues} (8.8\%), provide a clear direction for future studies. {\faLeanpub \hspace{0.5mm}} Researchers can delve deeper into these areas to develop innovative solutions that address architecture related limitations and challenges, thereby advancing the state of quantum software development. %\faMale \hspace{0.5mm} Practitioners understanding and mitigating identified challenges can lead to more efficient and reliable software development processes. 

\section{Threats to validity}\label{ThreatValidity}
\subsection{Construct validity}
Construct validity refers to whether our measurement tools accurately assess the intended properties \cite{ralph2018construct}. In this study, we encountered two potential challenges to construct validity: search term selection and manual analysis.
Search term selection: The retrieval of related issues from GitHub and related posts from Stack Exchange sites (including platforms such as Stack Overflow, Quantum Computing Stack Exchange, and Computer Science Stack Exchange) relies on keyword-based searches. When it came to selecting search keywords for finding GitHub issues and Stack Exchange posts, we faced challenges related to the construct validity of our data collection. One significant concern was whether the keywords we initially chose would capture all the relevant issues and posts or if some valuable information might be missed. To address this potential threat, we took a proactive approach by conducting pilot searches using alternative search terms before using the original terms discussed in Section \ref{datacollection}. This iterative process allowed us to refine our keyword choices, ensuring that we cast a wider net and minimize the risk of incomplete data, thus safeguarding the construct validity of our study.
Manual data handling: The stages of data handling in our study, which involve manual processes and thus risk subjectivity and bias, are as follows: (1) Labeling-- To minimize bias in this stage, we implemented a two-step approach. Initially, we performed a pilot labeling on both GitHub issues and Stack Exchange posts. This was done prior to the formal labeling to identify potential biases and inconsistencies. Furthermore, we established the inclusion and exclusion criteria to objectively decide which posts or issues should be included in our study. (2) Extraction: The pilot data extraction process was a collaborative effort primarily between the first and second authors. To ensure accuracy and consistency, the first author conducted a thorough recheck of the extraction results. This step was crucial to maintain the integrity of the data extraction process. If any discrepancies were found, the third author participated in discussions with the first two authors to help reach a consensus; (3) Analysis: To mitigate the threat of data analysis, we used two qualitative techniques (predefined classification and the Constant Comparison method) to analyze the extracted data and answer the RQs. Moreover, we attempted to minimize this threat by performing a pilot data analysis before the formal formal data analysis. The first and second authors independently categorized content from filtered posts and issues, cross-referencing the filtered posts and issues with the corresponding data items in Table \ref{dataExtraction}. The first author then consolidated these codes into higher-level concepts and categories, while the other authors reviewed and validated the pilot data analysis results, resolving disagreements through the negotiated agreement method \cite{campbell2013coding} to enhance reliability.

\subsection{External validity}
External validity encompasses the extent to which the outcomes of a study can be applied and verified in diverse settings, thereby assessing the generalization of the findings. One primary factor influencing external validity is the selection of data sources. In our effort to mitigate this potential limitation, we deliberately opted for two widely recognized developer communities, namely GitHub and Stack Exchange as the primary sources for extracting architecture decisions in the realm of quantum software development. It is important to note that GitHub and Stack Exchange have consistently served as pivotal resources in empirical software engineering investigations. Previous research endeavors (e.g., \cite{nasab2023empirical} and \cite{tian2019developers}) have demonstrated that insights obtained from GitHub and Stack Exchange data enjoy a high degree of validation and resonance among practitioners. This strong alignment between research findings and real-world experiences bolsters our confidence in the representatives of these platforms for studying architecture decisions in quantum software development. Furthermore, it is important to acknowledge that quantum computing is an emerging field still in its early stages compared to traditional software systems. As a result, the datasets available in this domain are often of limited sample size. For instance, Openja \textit{et al}. \cite{openja2022technical} conducted an empirical study on technical debt and faults in open-source quantum software systems using a dataset of 118 projects, highlighting the constraints and realities of research in this emerging field. This situation further emphasizes the importance of our study, which expands the initial dataset and continues to contribute to the understanding of quantum software architecture. The specific nature of our dataset reflects the current landscape of quantum software systems, which presents unique challenges for empirical research on this topic. We also acknowledge that encompassing more data sources would enhance the study’s external validity.

\subsection{Reliability}
Reliability, in the context of our research methodology, pertains to the degree of consistency exhibited by the chosen approach in yielding results. In order to mitigate the potential threat to reliability, a preliminary data labeling exercise was carried out independently by two authors on a sample consisting of 10 GitHub issues and 10 Stack Exchange posts. The assessment of inter-rater reliability was conducted through the calculation of Cohen's Kappa coefficient, yielding values of 0.80 for GitHub issues and 0.76 for Stack Exchange posts. Both of these coefficients exceed 0.7, indicating a satisfactory level of agreement between the participating authors. To further enhance the reliability of the process encompassing data labeling, data extraction, and data analysis, any disparities or inconsistencies that arose were resolved through constructive discussions involving the first author and other authors. To foster transparency and enable validation of our study findings, we have made the entire dataset, along with the corresponding data labeling outcomes for GitHub issues and Stack Exchange posts, readily available \cite{dataset}. This provision ensures that other researchers can scrutinize our work and verify the consistency and reliability of our data handling and analysis procedures.

\section{Related work}\label{relatedwork}
\subsection{Quantum software development}
In recent years, quantum software development has gained significant attention in the software engineering community, reflecting the growing importance of quantum computing. Khan \textit{et al}. \cite{khan2023software} conducted a systematic review to examine the software architecture for quantum computing systems. Their work provides a comprehensive overview of existing architectures of quantum software systems, highlighting the advancements and challenges in this domain. Truger \textit{et al}. \cite{truger2023warm} explored the intersection of warm-starting and quantum computing through a systematic mapping study, shedding light on how quantum computing can benefit from optimizing initialization procedures. Haghparast \textit{et al}. \cite{haghparast2023quantum} took a developer's perspective to identify the challenges in quantum software engineering and mapped these challenges to a proposed workflow model, offering insights into the practical issues developers face in this field. Yue \textit{et al}. \cite{yue2023challenges} examined the challenges and opportunities in quantum software architecture, highlighting key considerations in designing quantum computing software. Felderer \textit{et al}. \cite{felderer2023software} explored software engineering challenges in quantum computing through the First Working Seminar on Quantum Software Engineering (WSQSE 2022), providing valuable insights into this evolving domain. Finally, Zhao \textit{et al}. \cite{zhao2023bugs4q} made a contribution to the field of quantum programming by introducing Bugs4Q, which is a novel benchmark suite comprising forty-two real and manually validated Qiskit bugs extracted from prominent platforms such as GitHub, StackOverflow, and Stack Exchange. These bugs are supplemented with test cases that facilitate reproducing erroneous behaviors, and these bug scenarios are invaluable for systematically evaluating debugging and testing methodologies for quantum programs. These papers collectively contribute to the understanding of quantum software development, addressing various aspects such as software architecture, optimization, workflow models, challenges, and benchmarking in quantum computing systems.

\subsection{Decisions in software engineering}
Decisions are either explicitly presented in knowledge management tools or implied in various textual artifacts. Extensive research has been conducted to understand and analyze decisions and decision-making in software engineering. Li \textit{et al}. \cite{li2019decisions} conducted an exploratory study on the Hibernate developer mailing list, analyzing 9,006 posts to uncover insights into decision expressions, classification, rationale, approaches, related software artifacts, and the trend of decision-making over time. Their findings revealed that decisions in open-source software development are typically expressed as Information Giving, Solution Proposal, and Feature Request, with the main categories being Design Decision and Requirement Decision. In a separate study, Bi \textit{et al}. \cite{bi2018architecture} investigated how developers use architecture and design patterns concerning quality attributes and design contexts, revealing previously unknown relationships among these elements. Zhang \textit{et al}. \cite{zhang2023architecture} conducted an empirical study on architecture decisions in AI-based systems development. They analyzed the data from Stack Overflow and GitHub, revealing that architecture decisions in AI are expressed through various linguistic patterns, with Solution Proposal and Information Giving being common. Their study identified Technology Decision, Component Decision, and Data Decision as primary decision types, with a focus on Performance as the key quality attribute. It also highlighted Design Issues and Data Issues as significant challenges in the architecture decision-making of AI-based systems. Waseem \textit{et al}. \cite{waseem2022decision} introduced decision models aimed at assisting microservices practitioners in selecting appropriate patterns and strategies for Microservices Architecture (MSA). These models cover four key MSA design areas, including microservices application decomposition, security, communication, and service discovery. To create these models, the authors conducted an extensive literature review and evaluated their effectiveness through interviews with 24 practitioners, and the results show that these decision models were found to be valuable tools for guiding microservices patterns and strategy selection. Shahin \textit{et al}. \cite{shahin2010improving} explored the visualization of architectural design decisions, focusing on the use of the Compendium tool to improve understandability and communication in architectural design. Lastly, Liu \textit{et al}. \cite{liu2019understanding} provided insights into the decision-making of students in requirements engineering course projects, emphasizing the importance of understanding the characteristics of decision-making during various stages of requirements engineering. These works collectively contribute to our understanding of decision-making in software engineering, encompassing diverse facets such as decision expression, architecture patterns, visualization, and decision-making in the software development process.

\subsection{Architecture decisions in quantum software development}
Khan et al. \cite{khan2023software} conducted a systematic review to delve into the architectural aspects of quantum software development. Their study provides insights into how architectural processes, modeling notations, design patterns, tool support, and challenging factors impact quantum software architecture. The authors highlighted the transformation of quantum bits (Qubits) into architectural components, offering a fresh perspective on architectural decisions in quantum computing systems. Sodhi et al. \cite{sodhi2021quantum} investigated Quantum Computing Platforms (QCPs) from a software engineering perspective. They proposed a QCPs architecture and a programming model and evaluated the impact of QCPs characteristics on Quality Attributes (QAs) and Software Development Life Cycle (SDLC) activities. Their findings suggest that QCPs are best suited for specialized applications like scientific computing, with some QAs, such as maintainability and reliability, being negatively affected by certain QCPs characteristics (e.g., limitations in qubit state transmission, hard dependencies on quantum algorithms, and restrictions on copying and deletion of qubit states). Furthermore, Khan et al. \cite{khan2023agile} researched agile practices in quantum software development and delved into the challenges and applicability of agile practices in this domain. While their work primarily focuses on the adoption of agile practices, it also indirectly touches on limitations and challenges in quantum software development. Their study demonstrated that agile software development practices have the potential to address some of the challenges inherent in quantum software development, but it also highlighted new obstacles that impede their effective incorporation. This research offers valuable insights into the landscape of challenges faced when making architecture decisions in quantum software development.

Distinguished from the before mentioned research, our investigation delved into a comprehensive analysis of architectural decision-making within quantum software development. Specifically, our study encompassed an examination of the expressions and classifications of architecture decisions, the diverse application domains in which these decisions manifest, the critical quality attributes that factor into decision-making, as well as the formidable limitations and challenges encountered during this architectural process. To conduct this exploration, we conducted data mining across Stack Exchange and GitHub, amassing valuable insights into the intricacies of quantum software development's architectural landscape.

\section{Conclusions} \label{conclusionFurtureWork}
Quantum software architecture refers to the design and organization blueprint for quantum software systems that leverage the principles of quantum computing to solve specific problems or tasks. It focuses on addressing the challenges and limitations inherent to quantum computation, providing a structured approach to building applications that leverage the power of quantum technologies. Architecture decisions are essential for architects and developers to make informed design choices during the development of quantum software systems, and once a system is developed, it can be challenging and costly to change at the architecture level. This research aimed to explore the architecture decisions within the context of quantum software development. Specifically, we conducted an empirical study on GitHub issues and Stack Exchange posts to explore architecture decisions in quantum software development from practitioners' perspectives. We used a keyword-based search to collect data from GitHub and Stack Exchange. Finally, we got 385 GitHub issues and 70 Stack Exchange posts which include 33 Stack Overflow posts, 33, Quantum Computing Stack Exchange posts, and 4 Computer Science Stack Exchange posts. We manually extracted architecture decisions related to quantum software systems to investigate decision expressions, decision types, involved application domains, quality attributes considered, and limitations and challenges encountered in architecture decision-making of quantum software development. The main results are that:

\begin{itemize}
 \item In quantum software development, architecture decisions are expressed through six distinct linguistic patterns, with \textit{Solution Proposal} and \textit{Information Giving} being the most prevalent. \textit{Solution Proposal} constitutes 35.83\%, while \textit{Information Giving} accounts for 32.80\% of the linguistic patterns used. Conversely, \textit{Opinion Asking} and \textit{Problem Discovery} are relatively less used in the linguistic patterns we collected.
\item Architecture decisions primarily fall into two major categories: \textit{Implementation Decision} and \textit{Technology Decision}. Specifically, \textit{Implementation Decision} represents 22.10\%, and \textit{Technology Decision} accounts for 18.81\% of architecture decisions we collected.
\item\textit{Software Development Tools} stand out as the most common application domain among the twenty domains identified, representing 22.15\% of the application domain we collected.
\item Two important quality attributes considered in architectural decision-making of quantum software systems are \textit{Maintainability} (18.53\%) and \textit{Performance} (15.68\%) of the quality attributes we collected.
\item Practitioners in quantum software development face significant limitations and challenges. \textit{Design Issues} are prominent, accounting for 25.07\%, while \textit{High Error Rates} are a notable concern at 11.37\% of the limitations and challenges we collected.
\end{itemize}

The findings of this study offer valuable insights to researchers and practitioners. By identifying common linguistic patterns, major decision categories, prevalent application domains, and critical quality attributes, this research equips professionals in quantum software systems with a deeper understanding of the made architectural decisions. Additionally, it highlights common challenges of making architecture decisions in quantum software development, such as design and performance issues, providing researchers and practitioners with the awareness needed to address these issues. 

In the next step: (1) We intend to conduct a comparative study of architecture decisions on GitHub, Stack Exchange, and other platforms like developer mailing lists, aiming to reveal insights into the current focus of architecture decision-making in quantum software systems, including its advantages and deficiencies. (2) To provide a comprehensive perspective, we will validate and extend the proposed taxonomy of architecture decisions using a practitioner survey or interview, collecting insights from professionals actively engaged in quantum software development. (3) Finally, our overarching goal is to address the limitations and challenges identified in this study. We aim to develop decision models for selecting architecture patterns and strategies in architecting quantum software systems, contributing to more informed and effective architecture decision-making in quantum software development.

\section*{\textbf{Data availability}}
We have shared the link to our dataset in the reference~\cite{dataset}.

\section*{\textbf{Acknowledgements}}
This work has been partially sponsored by the National Natural Science Foundation of China (NSFC) under Grant No. 62172311 and the Special Fund of Hubei Luojia Laboratory. The authors would also like to acknowledge the financial support from the China Scholarship Council.

\bibliography{mybibfile}
\end{sloppypar}
\end{document}